\documentclass[apj]{emulateapj}

\newcommand{\myemail}{zhao@cfa.harvard.edu}
\newcommand{\cpl}{{\it ChaMPlane }}
\slugcomment{To appear in ApJ Supplement}

\shorttitle{\cpl Survey: Photometry}
\shortauthors{Zhao et al.}

\newcommand{\figmosaicgal}{
\begin{figure}
\centering\includegraphics[trim=130 80 130 70,scale=0.4,angle=-90]
{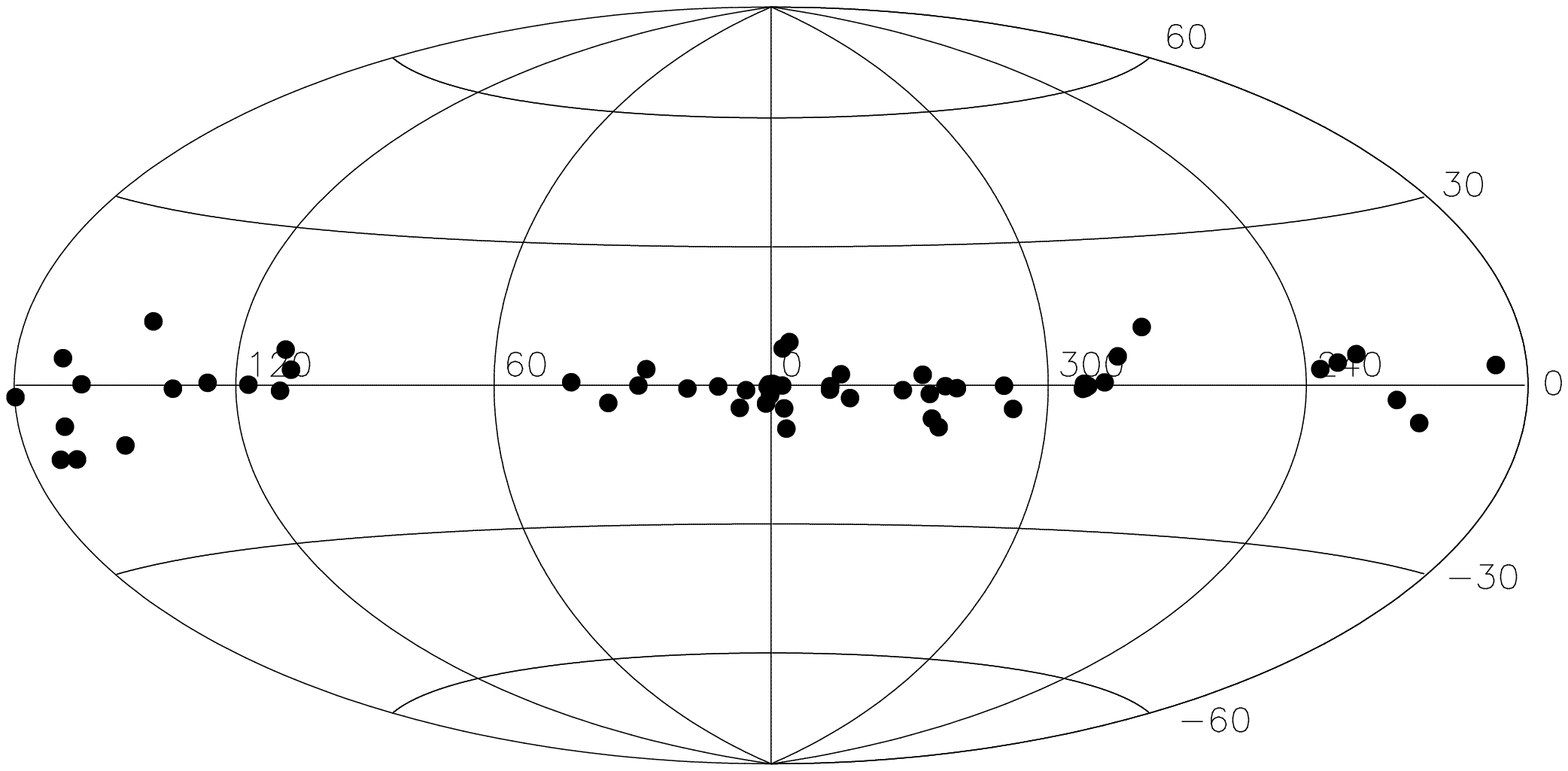}
\caption{\label{fig:mosaicgal}Observed \cpl Mosaic fields in Galactic
coordinates.  Several fields in the Galactic center region are
unresolved on this figure (see \citet{gri05} for an expanded
view of the galactic center region fields.).}
\end{figure}
}

\newcommand{\figastrometry}{
\begin{figure}
\centering\includegraphics*[trim=0 10 0 0,width=3.5in]{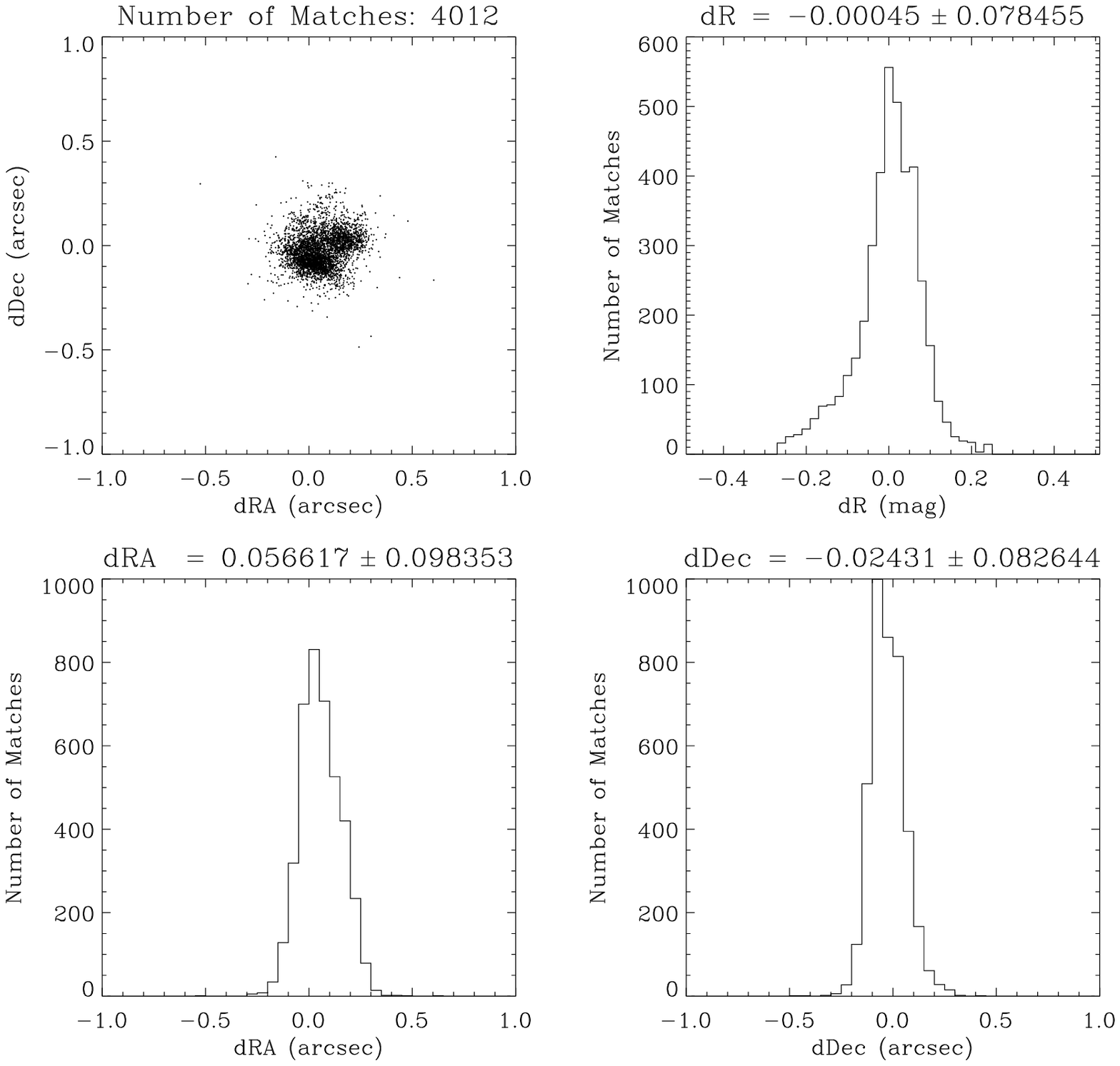}
\caption{\label{fig:astrometry}The astrometric precision of the Mosaic
photometry.  The upper left panel shows the offset of the same stars
between the GC1 and GC2 fields in their overlapping area.  The upper
right panel is a histogram showing the R magnitude difference of the
same stars between the GC1 and GC2 optical catalogs.  The two lower
panels are the histograms of the offset in RA and Dec of these stars.}
\end{figure}
}

\newcommand{\figsnrfactor}{
\begin{figure}
\centering\includegraphics*[width=3in]{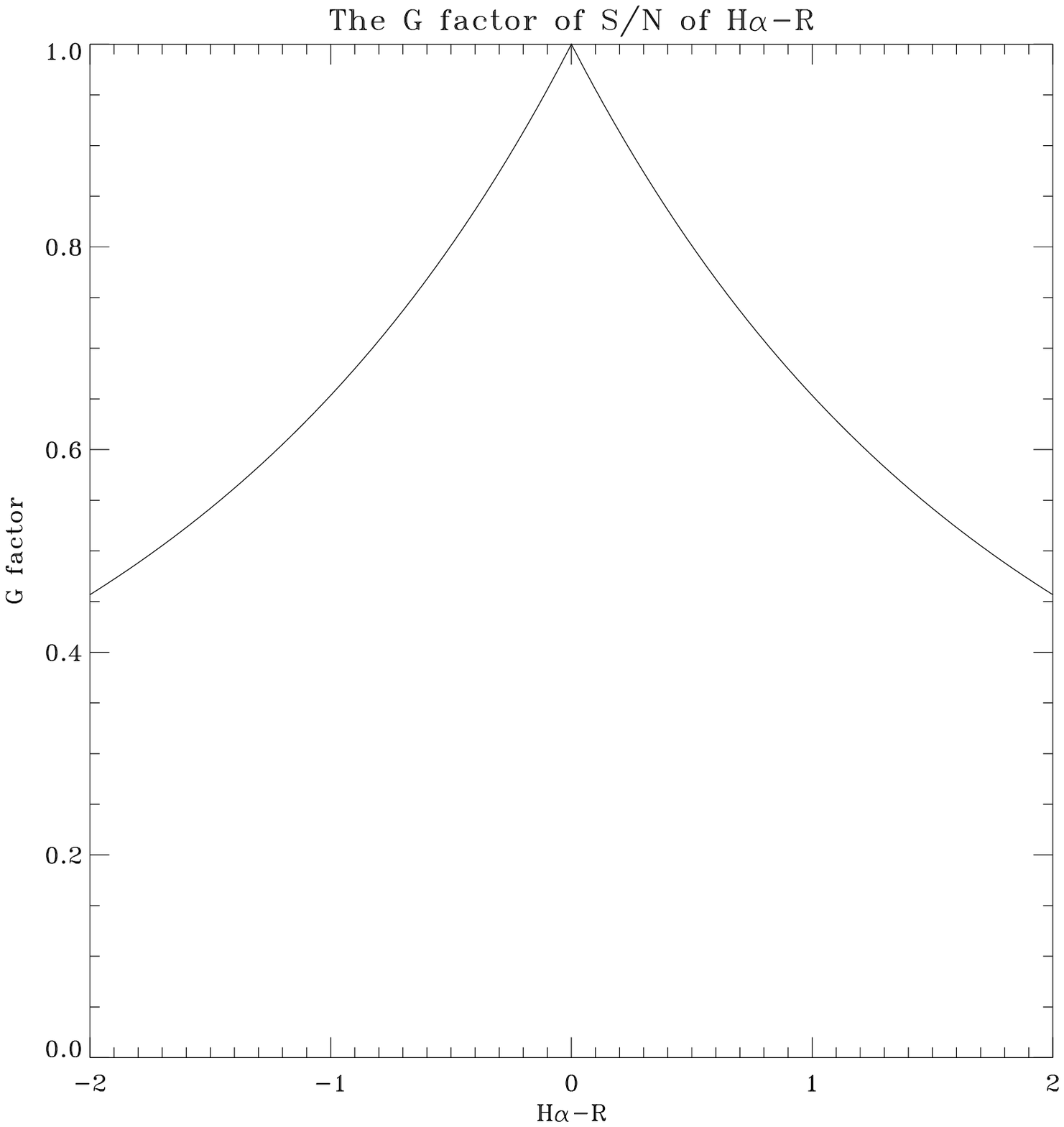}
\caption{\label{fig:snrfactor} The G factor of signal-to-noise ratio,
{\em S/N}, in terms of the flux ratio, {\em F}, between H$\alpha$ and
R bands.}
\end{figure}
}

\newcommand{\fighaew}{
\begin{figure}
\centering\includegraphics*[trim=0 3 0 0,width=3.0in]{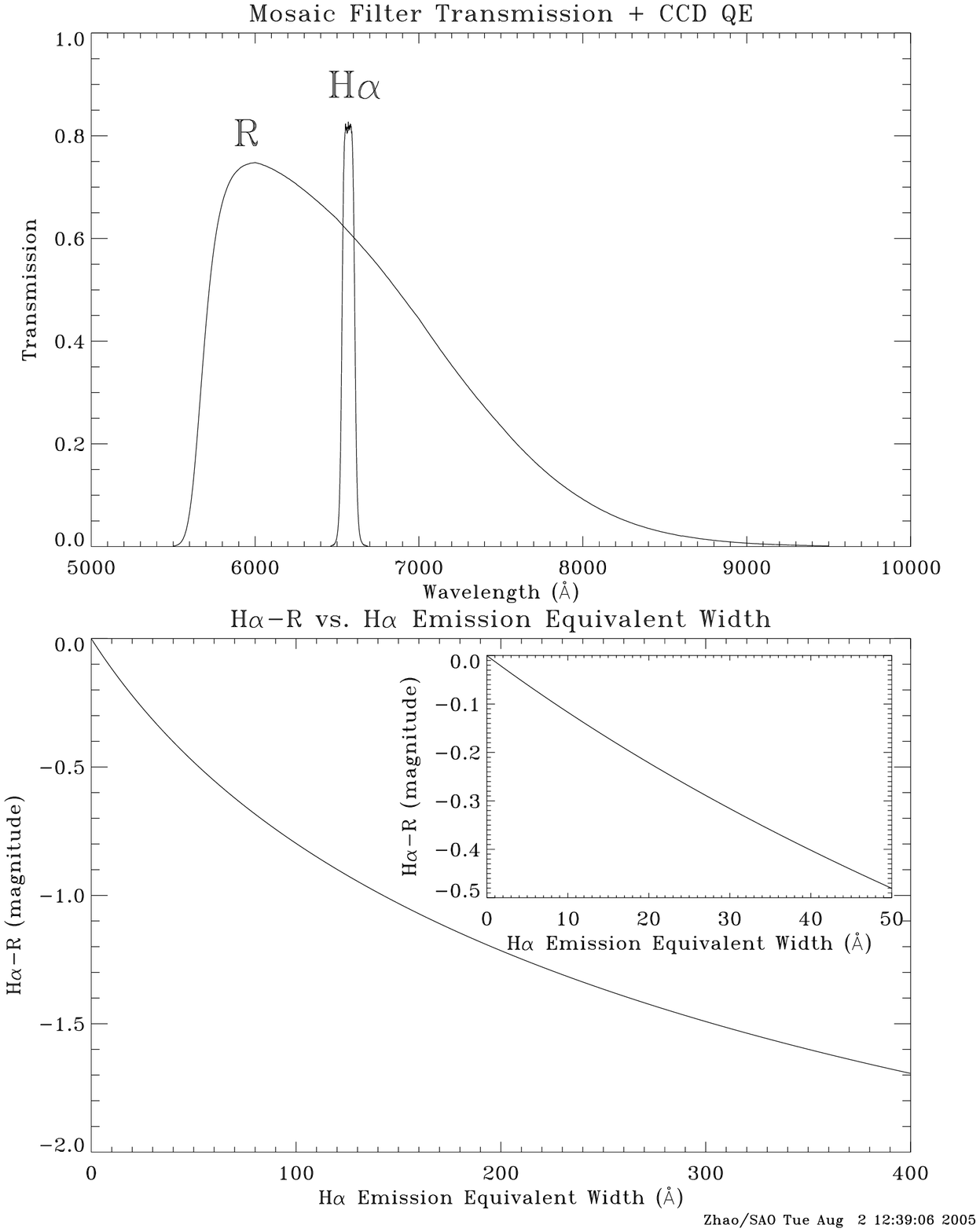}
\caption{\label{fig:haew} Top panel: the composite transmission curves
  of the Mosaic R and H$\alpha$ filters plus the CCD quantum
  efficiency.  Bottom panel: H$\alpha-$R as a function of the
  H$\alpha$ emission equivalent width (assuming a flat continuum).}
\end{figure}
}

\newcommand{\fighaewms}{
\begin{figure}
\centering\includegraphics*[trim=0 0 0 0,width=3.2in]{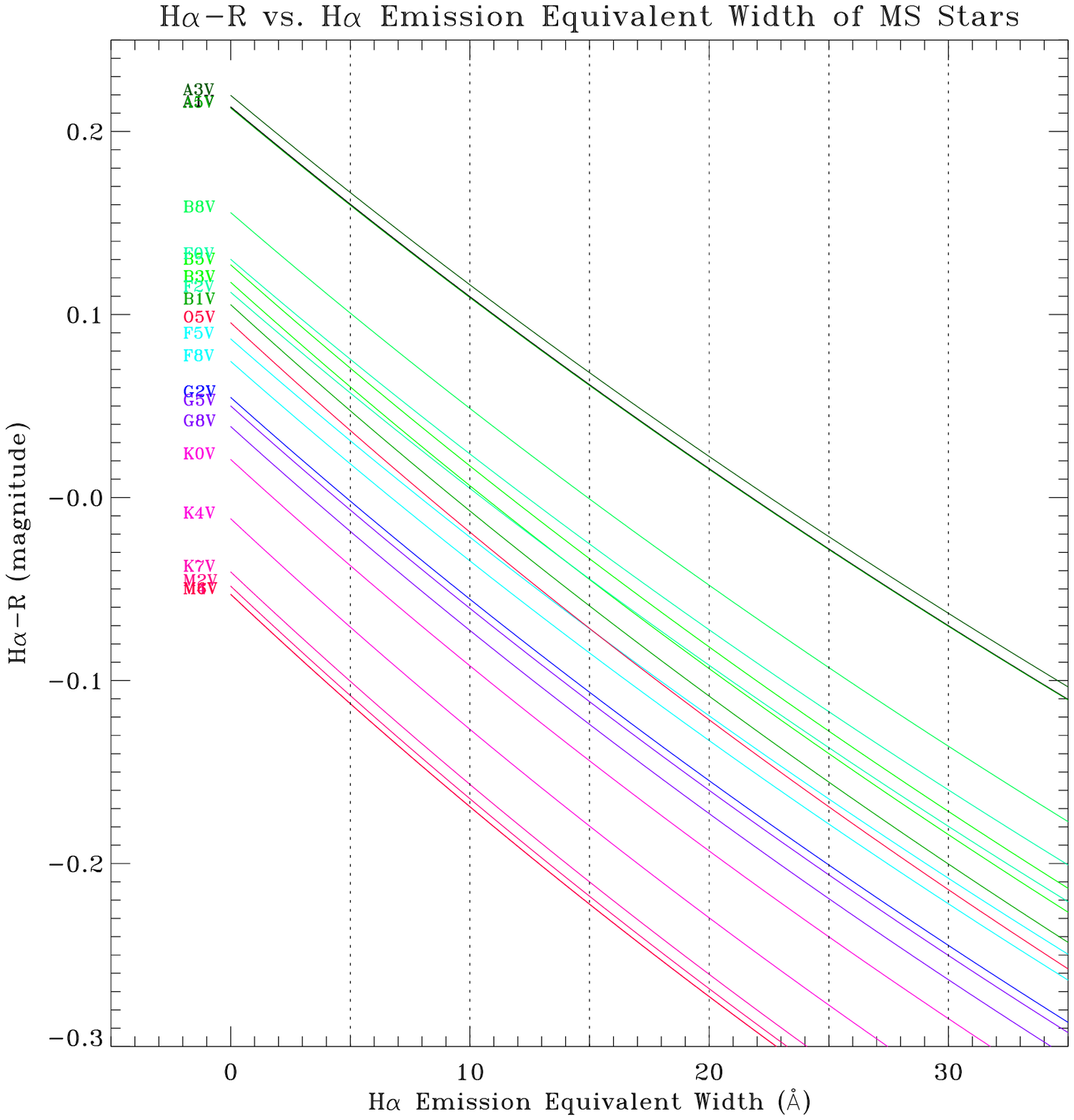}
\caption{\label{fig:haewms} H$\alpha-$R as a function of the H$\alpha$
emission equivalent width of main sequence stars for the Mosaic H$\alpha$
and R filters.  (See electronic ApJ for the colored version.)}
\end{figure}
}

\newcommand{\fighasources}{
\begin{figure}
\centerline{\LARGE ~R \hspace{1.4in}H$\alpha$}

\centering\includegraphics*[trim=0 0 0 0,scale=0.45,angle=0]
{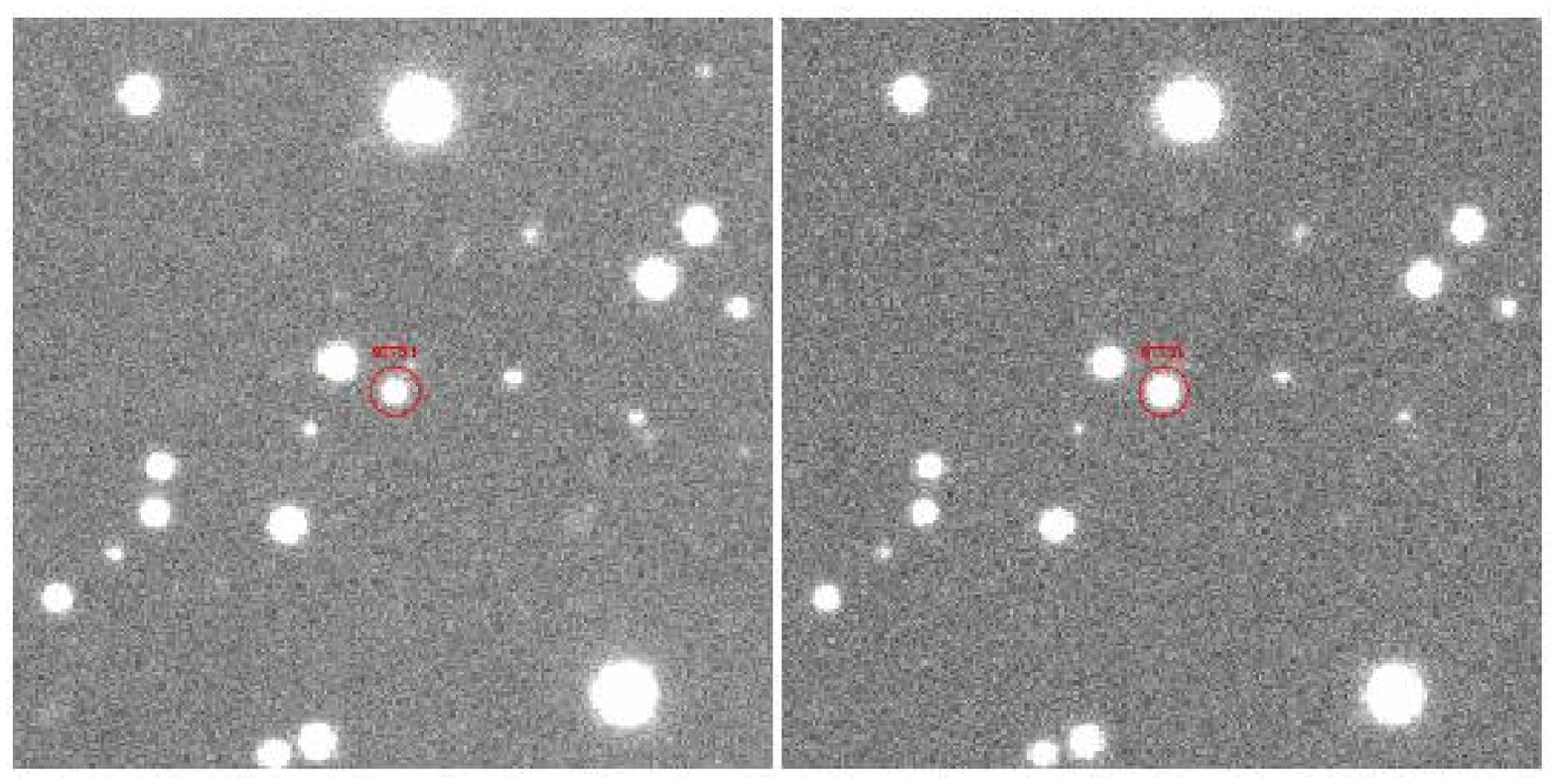}
\centering\includegraphics*[trim=0 0 0 0,scale=0.45,angle=0]
{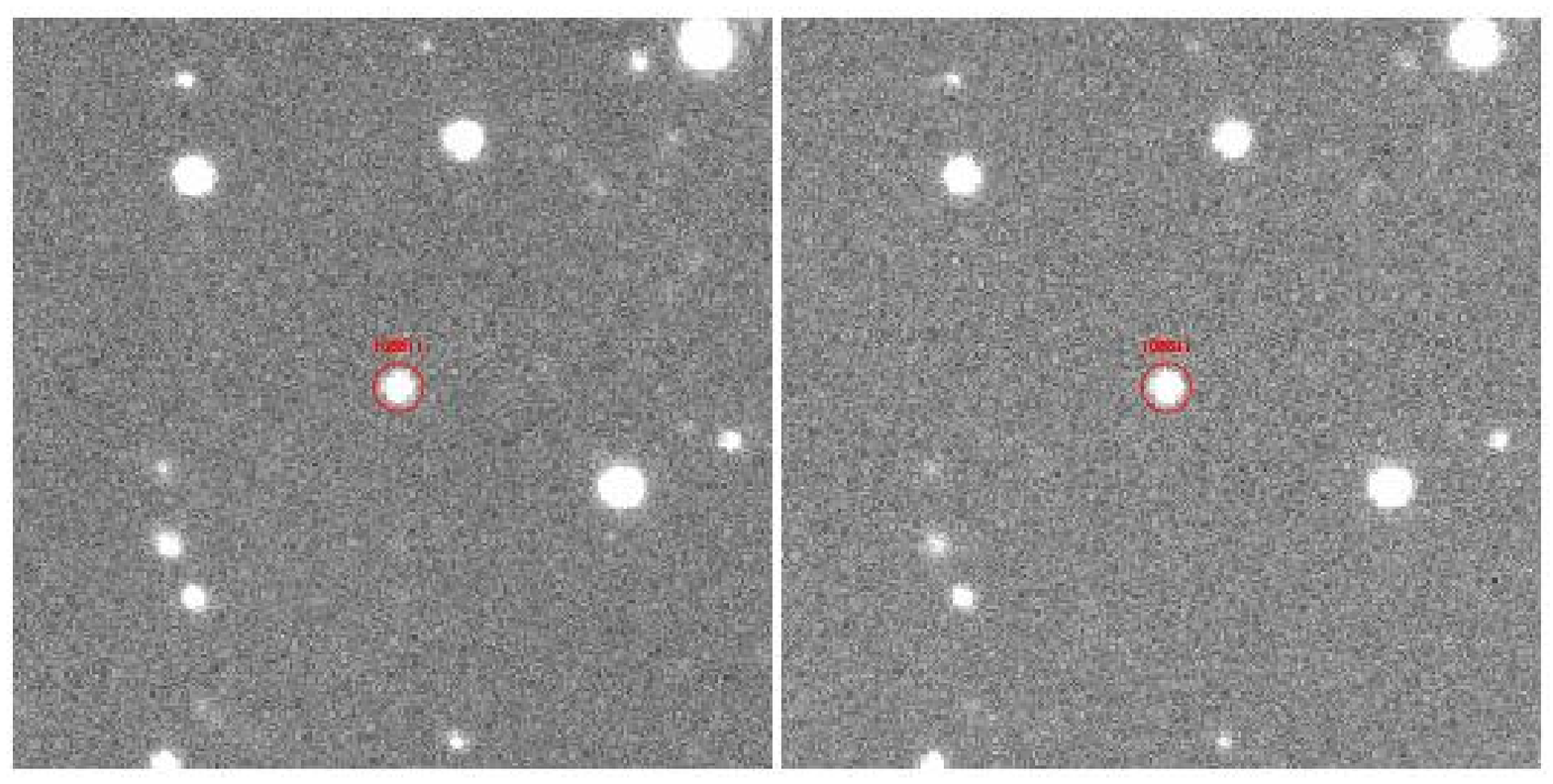}
\centering\includegraphics*[trim=0 0 0 0,scale=0.45,angle=0]
{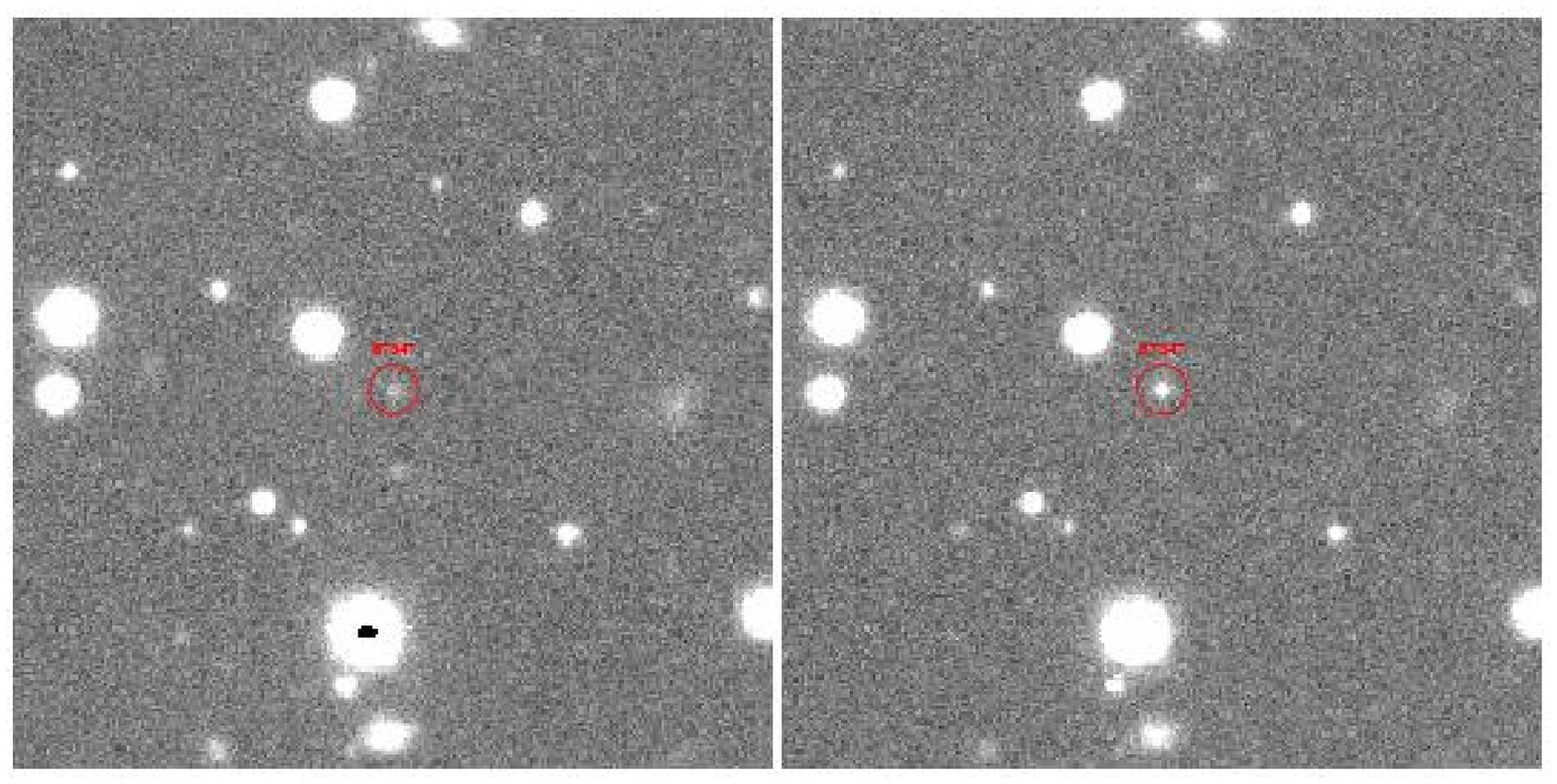}
\caption{\label{fig:hasources}The R (left) and H$\alpha$ (right) 
  images of three H$\alpha$ emission sources found in the J0422+32
  Mosaic field: 1) top panels show the GRO J0422+32 itself (circled),
  which is obviously brighter in H$\alpha$ (ID=97731, R~=~20.80,
  H$\alpha-$R~=~$-$1.45); 2) middle panels show the first CV
  discovered under \cpl project, which is outside of the ACIS FoV
  (ID=108811, R~=~20.28, H$\alpha-$R~=~$-$0.76); 3) bottom panels show
  a very bright H$\alpha$ emission object (ID=87347, R~=~21.67,
  H$\alpha-$R~=~$-$1.54).  It is in the ACIS-I FoV.  Although no X-ray
  emission was detected from this object, the upper limit for
  $F_x/F_r$ allows it to be a CV.  ($1\arcmin\times1\arcmin$ FoV.)}
\end{figure}
}

\newcommand{\figJxoidha}{
\begin{figure*}
\centering\includegraphics*[trim=0 0 0 0,scale=0.85,angle=0]
{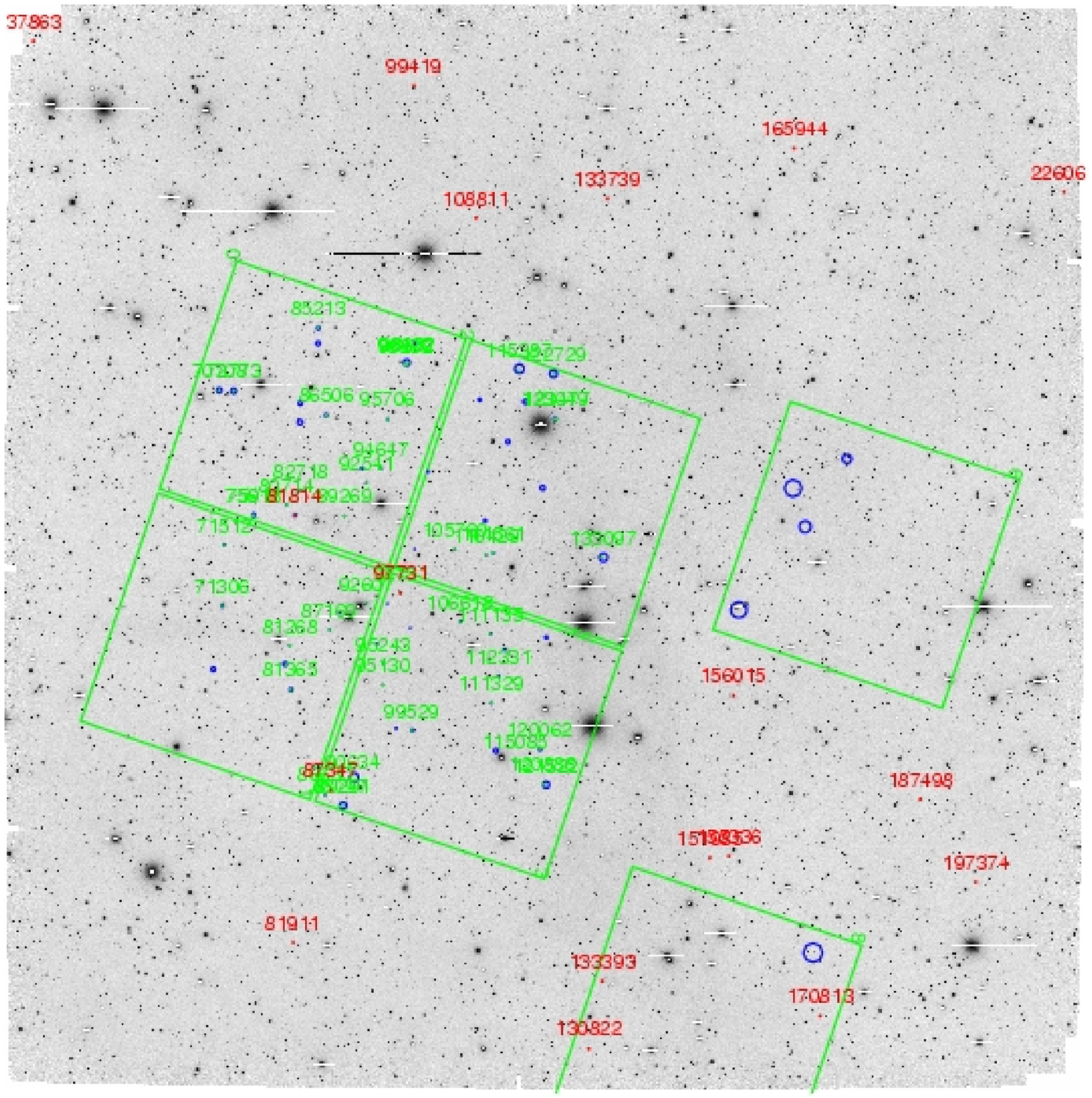}
\caption{\label{fig:J0422_x_o_id_ha}GRO J0422+32 field full Mosaic deep R
  image with active ACIS CCDs overlay (large green squares, ACIS I0-3
  and S2,4 were turned on).  Chandra source 3-$\sigma$ error circles
  are marked with blue color.  Chandra candidate optical counterparts
  are marked by small (1\arcsec) green circles with their optical ID.
  H$\alpha$ emission sources (H$\alpha -$R$\leqslant$$-0.3$ and
  S/N$\geqslant$5) are marked with (1.5\arcsec) red circles with their
  optical ID.  ID~108811 (above the ACIS-I) is the first CV discovered
  under the \cpl project.  ID~87347 (near the bottom of ACIS-I) is the
  previously unknown bright H$\alpha$ emission object without X-ray
  detection.  Five X-ray sources are detected on the ACIS-S2 and S4
  chips.  We do not include their optical counterparts here as their
  positional errors are relatively poorly known due to their large
  off-axis angles.  (North-up, east-left; $36\arcmin\times36\arcmin$
  FoV.)  (See electronic ApJ for the colored version.)}
\end{figure*}
}

\newcommand{\figJxoidaimha}{
\begin{figure*}
\centering\includegraphics*[trim=0 0 0 0,scale=0.85,angle=0]
{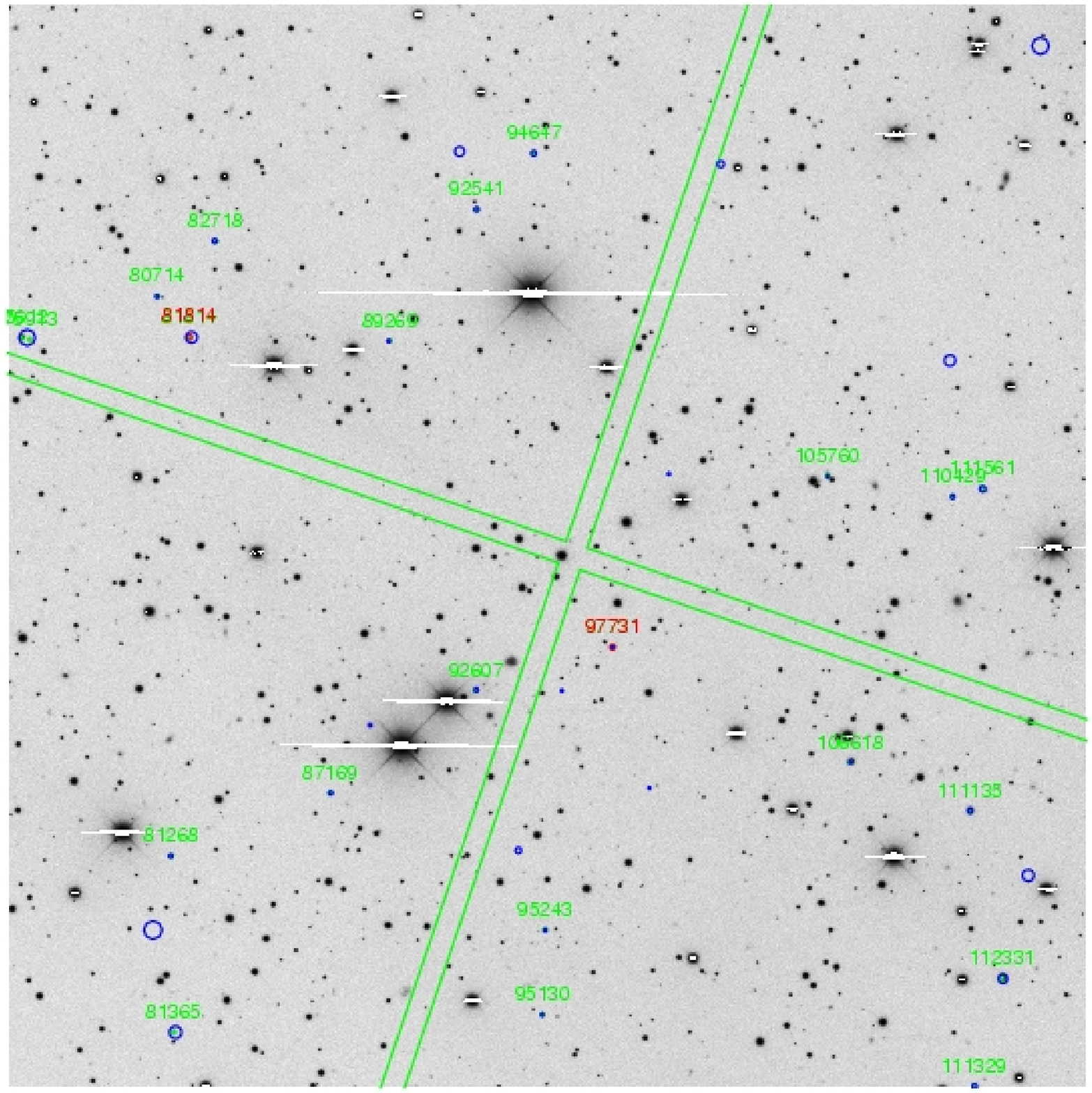}
\caption{\label{fig:J0422_x_o_id_aim_ha}GRO J0422+32 field Mosaic deep
  R image around the ACIS-I aimpoint (See Figure
  \ref{fig:J0422_x_o_id_ha} for source marker notations).  ID~97731 at
  the aimpoint is the target J0422+32.  ID~81814 is the only other
  H$\alpha$ emission Chandra source on ACIS-I, which is a QSO at
  z=4.25, with Lyman-$\alpha$ red-shifted to Balmer-$\alpha$
  \citep{rog05}.  (North-up, east-left.  ($9\arcmin\times9\arcmin$
  FoV.)  (See electronic ApJ for the colored version.)}
\end{figure*}
}

\newcommand{\figJcmdxmacis}{
\begin{figure*}
\centering\includegraphics*[trim=0 10 0 0,scale=0.85,angle=0]
{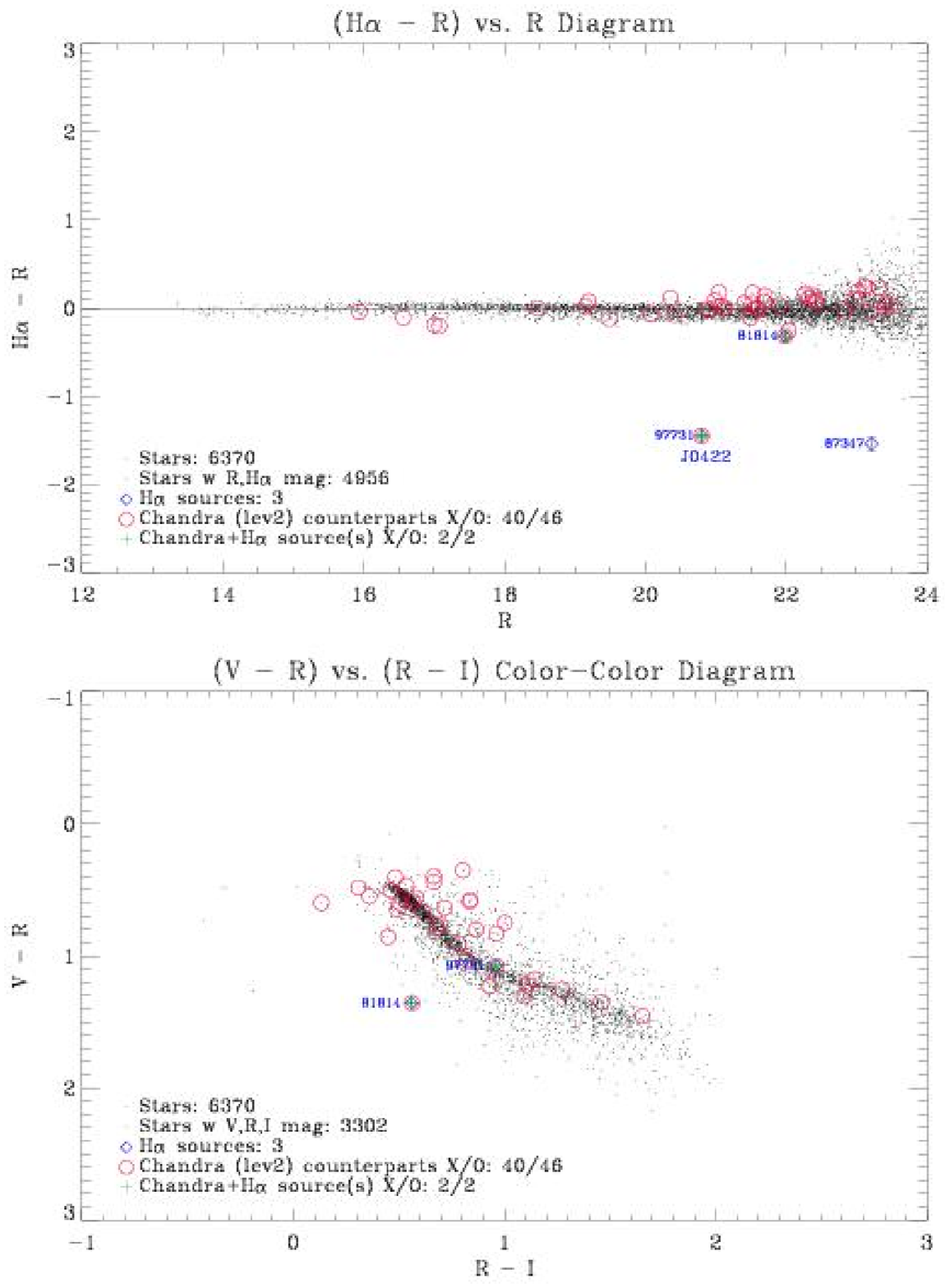}
\caption{\label{fig:J0422_cmd_xm_acis}GRO J0422+32 field (H$\alpha-$R)
vs. R color-magnitude and (V$-$R) vs. (R$-$I) color-color diagrams of
objects inside ACIS-I FoV.  40 Chandra sources are matched with 46
counterparts (marked as red circles).  There are three H$\alpha$
emission sources (marked as blue diamonds) within the ACIS-I FoV.  Two
of them are Chandra sources: ID~97731 is the black-hole X-ray nova
J0422+32 ; ID~81814 is a quasar at z=4.25.  ID~87347 is an unclassified
bright H$\alpha$ emission object without X-ray detection.  (See
electronic ApJ for the colored version.)}
\end{figure*}
}

\newcommand{\figJcmdxmout}{
\begin{figure*}
\centering\includegraphics*[trim=0 10 0 0,scale=0.85,angle=0]
{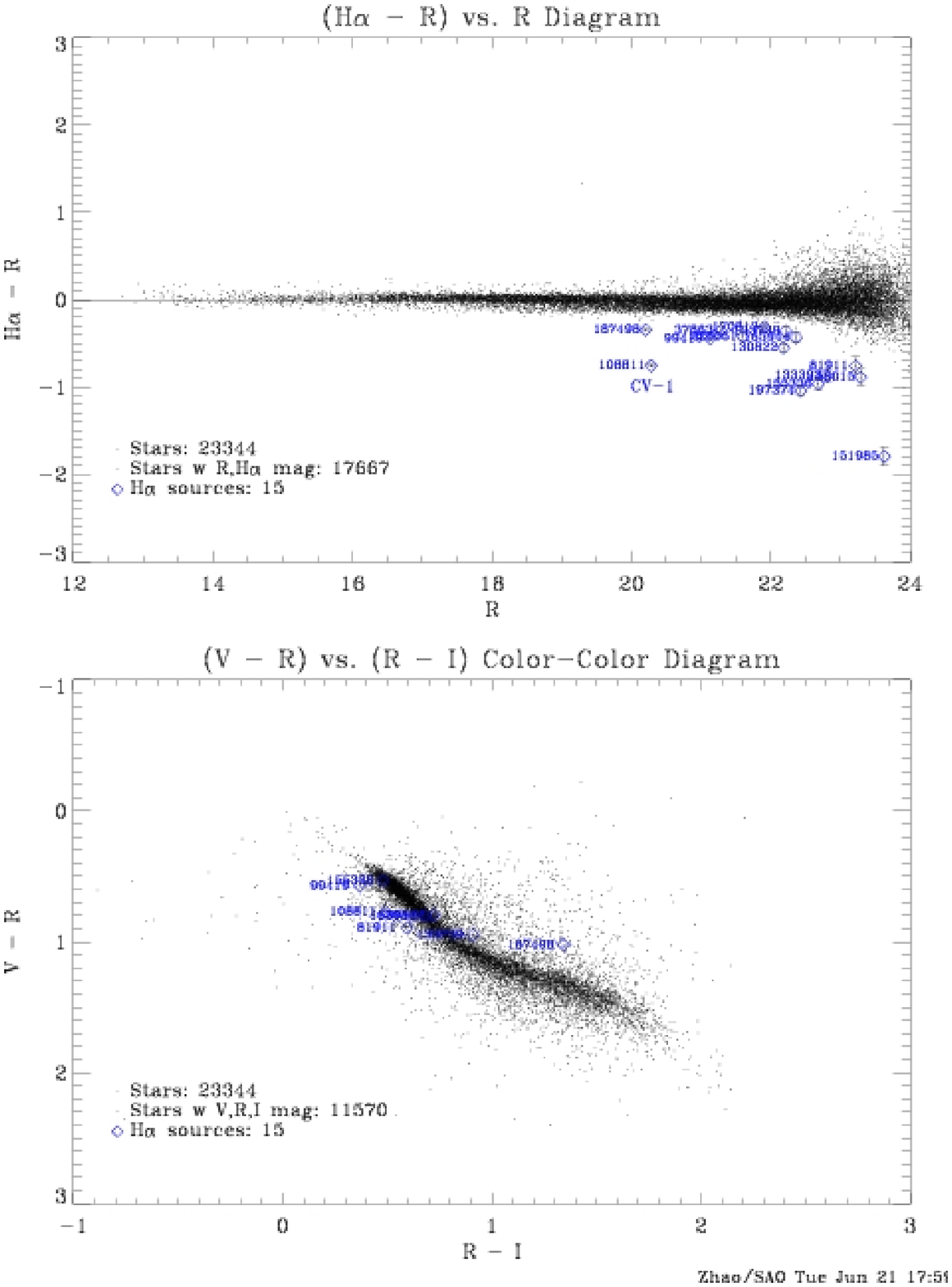}
\caption{\label{fig:J0422_cmd_xm_out}GRO J0422+32 field (H$\alpha -$R)
vs. R color-magnitude and (V$-$R) vs. (R$-$I) color-color diagrams of
objects outside ACIS-I FoV.  There are 15 H$\alpha$ emission sources.
The object with ID~108811 is the first CV discovered under the \cpl
project.  (See electronic ApJ for the colored version.)} 
\end{figure*}
}

\begin{document}

\title{ChaMPlane Optical Survey: Mosaic Photometry}

\author{Ping Zhao\altaffilmark{1,2,*}, Jonathan E. Grindlay,
Jae Sub Hong, Silas Laycock\altaffilmark{1,2}, Xavier P.
Koenig\altaffilmark{1}, Eric M. Schlegel, and Maureen van den
Berg\altaffilmark{2}}
\affil{Harvard-Smithsonian Center for Astrophysics, 60 Garden Street,
  Cambridge, MA 02138 USA}
%\email{\myemail}

\altaffiltext{*}{Send request to Ping Zhao at \myemail}
\altaffiltext{1}{Visiting Astronomer, Cerro Tololo Inter-American
  Observatory.  National Optical Astronomy Observatory, which is
  operated by the Association of Universities for Research in
  Astronomy, Inc.\ (AURA) under cooperative agreement with the
  National Science Foundation.}
\altaffiltext{2}{Visiting Astronomer, Kitt Peak National Observatory,
  National Optical Astronomy Observatory, which is operated by the
  Association of Universities for Research in Astronomy, Inc.\ (AURA)
  under cooperative agreement with the National Science Foundation. }

\begin{abstract}
The \emph{ChaMPlane} survey to identify and analyze the serendipitous
X-ray sources in deep Galactic plane fields incorporates the \cpl
Optical Survey, which is one of NOAO's Long-term Survey Programs.  We
started this optical imaging survey in March 2000 and completed it in
June 2005. Using the NOAO 4-m telescopes with the Mosaic cameras at
CTIO and KPNO, deep images of the \cpl fields are obtained in V, R, I
and H$\alpha$ bands.  This paper describes the process of observation,
data reduction and analysis of fields included in the \cpl Optical
Survey, and describes the search for H$\alpha$ emission objects and
Chandra optical counterparts.  We illustrate these procedures using
the \cpl field for the black hole X-ray binary GRO J0422+32 as an
example.
\end{abstract}


\section{Introduction}
\label{sec:introduction}
\setcounter{footnote}{2}

The \cpl Survey\footnote{\label{cplweb}http://hea-www.harvard.edu/ChaMPlane}
identifies serendipitous X-ray sources located to arcsec precision in
the Galactic plane fields from the Chandra archive, in order to
determine the populations of accretion-powered binaries in the Galaxy
(\citet{gri05}, and see \citet{gri02b, zha02b} for early descriptions).

The primary goals of this survey are: 1) to identify Cataclysmic
Variables (CVs) and quiescent Low Mass X-ray Binaries (qLMXBs: which
contain either black hole or neutron star primaries) in order to
measure their number and space density luminosity functions; and 2) to
determine the distributions of High Mass X-ray Binaries with Be star
secondaries.  The secondary goal is to study the distributions of
stellar coronal source and diffuse X-ray objects in the Galactic
Plane.  See \citet{gri05} for a complete description of the survey
goals, selection criteria and initial results.

\cpl consists of an X-ray and an Optical survey.  The X-ray survey,
supported by NASA through Chandra archival proposals, searches for
X-ray sources detected serendipitously in Chandra archival fields.  In
the Optical Survey, supported by NOAO, we look for H$\alpha$ emission
sources and the Chandra optical counterparts in 4-meter telescope
Mosaic images taken at CTIO and KPNO.  We successfully conducted the
\cpl Optical Survey from March 2000 to June 2005.  This 5 year survey
produced 65 Mosaic fields covering about 23 square degrees and 154
ACIS observations on 105 distinct Chandra fields (defined as groups of
individual observations whose aimpoints are $\geqslant$4\arcmin\
apart) in the Galactic plane during Chandra Cycles 1-6.

In this paper we describe the methods of the \cpl Optical Survey and
the procedures to search for H$\alpha$ emission objects and Chandra
optical counterparts.  Section \ref{sec:imaging} describes the
observational approach.  Section \ref{sec:reduction} describes the
Mosaic data reduction.  Section \ref{sec:photometry} describes the
photometric analysis and calibration.  Section \ref{sec:opticalcat}
describes the photometric results -- the optical catalog.  Section
\ref{sec:astrometry} gives the astrometric accuracy of the catalog.
Sections \ref{sec:ha} and \ref{sec:chandra_opt} describe the search
for H$\alpha$ emission objects and Chandra optical counterparts,
respectively.  Section \ref{sec:results} describes the data product,
and uses the results from the GRO J0422+32 field as an example.

\section{Optical Imaging}
\label{sec:imaging}

%\figmosaicgal 

Under the NOAO Long-term Survey Program, we were granted 5 nights CTIO
4-m and 1-2 nights KPNO 4-m telescope time each year for 5 years.
Before this long term program officially started in December 2000 at
KPNO, we conducted a pilot run using the CTIO 4-m Mosaic on March
13-15, 2000.  Under this Survey Program, we are committed to establish
an archival database to provide the community with all of our optical
images as well as a photometrically-calibrated star catalog.

\cpl fields are selected from the Chandra target list based on
criteria given in \citet{gri05}.  Deep optical imaging is the crucial
first step of the \cpl survey.  It serves two purposes: 1) to identify
candidate optical counterparts of the Chandra sources and to measure
their optical magnitudes and $F_X/F_{opt}$ ratios for approximate
spectral classification and constraints on reddening; and 2) to
identify CVs and qLMXBs by their ubiquitous H$\alpha$ excess as
``blue'' objects in the R vs. (H$\alpha -$ R) diagram.  It also paves
the way for the next step -- spectroscopic follow-up for
classification on identified \cpl objects.

\subsection{Instrument}

The images were taken with the Mosaic-I (KPNO) and Mosaic-II (CTIO)
cameras\footnote{The complete instrument information can be found at
{\tt http://www.noao.edu/noao/mosaic/}}.  The Mosaic camera is a
8k$\times$8k CCD array, which consists of eight 2048$\times$4096 SITe
CCDs.  The pixel size is 15~$\mu$m (0.26$\arcsec$ at the 4-m
telescope), which gives adequate resolution.  It provides a field of
view of 36$\arcmin\times$36$\arcmin$, which covers the full
Chandra/ACIS FoV.  Images were taken with filters of Johnson V, R, I
and H$\alpha$ (80\AA\ FWHM centered on the H$\alpha$ line).

To prepare the observation, the center of each Mosaic field (i.e. the
telescope pointing) was carefully positioned so that the Mosaic covers
all the active ACIS chips (typically, six ACIS chips are turned on).
For the Mosaic fields covering multiple Chandra observations, their
centers were positioned so that the Mosaic can cover maximum numbers
of active ACIS chips possible.

\subsection{Observations}
\label{subsec:onservations}

Table \ref{tbl_phot_obs} lists all the \cpl Optical Imaging runs
conducted.  Table \ref{tbl_mosaic} is a complete list of all 65
observed \cpl Mosaic fields.  The name of each field is based on the
main Chandra observation covered by that Mosaic field, except for
fields GC1 -- GC6, which are our Galactic Center mapping of 2.2 square
degrees that covers 58 Chandra ACIS observations in Cycles 1-6
(including 20 SgrA* \citep{mun03}, 30 Galactic Center Survey
\citep{wan02} and 8 other observations in the Galactic Center region).
Figure \ref{fig:mosaicgal} shows these 65 fields in Galactic
coordinates.  \begin{table}
\begin{center}
\caption{\label{tbl_phot_obs}ChaMPlane Observations: Mosaic Imaging}

\begin{tabular}{llrc} \hline \hline
Code   & Telescope       & Observing Date\tablenotemark{*} & Fields \\ \hline
ctio00 & CTIO 4-m & Mar. 13 -- 15, 2000 & 9 \\
kpno00 & KPNO 4-m & Dec.  5 -- ~6, 2000 & 4 \\
ctio01 & CTIO 4-m & May  13 -- 17, 2001 & 8  \\
kpno01 & KPNO 4-m & Oct. 25, 2001       & 3  \\
kpno02 & KPNO 4-m & Dec.  7 -- ~8, 2002 & 6 \\
ctio03 & CTIO 4-m & Jun.  2 -- ~6, 2003 & 11 \\
kpno03 & KPNO 4-m & Jan. 30 -- 31, 2004 & 2 \\
ctio04 & CTIO 4-m & May  16 -- 20, 2004 & 13 \\
kpno04 & KPNO 4-m & Jan. 11 -- 12, 2005 & 3 \\ 
ctio05 & CTIO 4-m & Jun. 7 -- 11, 2005 & 6 \\ \hline
Total &&&  65 \\ \hline \hline
\end{tabular}
\tablenotetext{*}{CTIO observations in 2002 were completely clouded.}
\end{center}
\end{table}
 % column N Mosaic MosaicField RAm Decm lm bm NHe22m </data/champ3/champlane/summary/mosaic.rdb > tbl_mosaic.rdb
% rdb2latex tbl_mosaic.rdb > tbl_mosaic
\begin{table*}
%\small
%\footnotesize
%\scriptsize
\begin{center}
\caption{\label{tbl_mosaic} 65 \cpl Mosaic Fields (sorted by RA)}
\begin{tabular}{rclccrrr} \hline \hline
N\tablenotemark{a} & Code   & Field & RA(J2000) & Dec(J2000) & $l~(^{\circ})$~~ & $b~(^{\circ})$~~ & N$_{\rm H}/10^{22}$\tablenotemark{b} \\ \hline
26 & kpno02 & G127.1$+$0.5 & 01:28:00.00 & $+$63:03:11.7 & 127.06023 & 0.47507 & 0.869 \\
58 & kpno04 & MAFFEI1 & 02:36:49.80 & $+$59:36:19.0 & 135.90926 & $-$0.58435 & 0.591 \\
11 & kpno00 & GKPERSEI & 03:30:37.89 & $+$43:51:15.8 & 150.90053 & $-$10.20409 & 0.181 \\
12 & kpno00 & GROJ0422$+$32 & 04:21:20.40 & $+$32:55:37.9 & 165.81091 & $-$11.95536 & 0.189 \\
13 & kpno00 & NGC1569 & 04:30:25.00 & $+$64:44:53.0 & 143.72842 & 11.14309 & 0.412 \\
23 & kpno01 & 3C123 & 04:36:28.00 & $+$29:37:49.0 & 170.52481 & $-$11.78690 & 0.659 \\
42 & kpno03 & AFGL618 & 04:42:23.39 & $+$36:04:35.3 & 166.40825 & $-$6.62965 & 0.462 \\
24 & kpno01 & 3C129 & 04:49:19.20 & $+$45:02:30.0 & 160.42311 & 0.18431 & 0.644 \\
27 & kpno02 & G166.0$+$4.2 & 05:27:12.06 & $+$42:54:05.1 & 166.21402 & 4.36698 & 0.378 \\
28 & kpno02 & PSRJ0538$+$2817 & 05:38:00.54 & $+$28:14:24.0 & 179.70937 & $-$1.78641 & 0.759 \\
43 & kpno03 & 1SAXJ0618.0$+$2227 & 06:18:18.70 & $+$22:29:20.8 & 189.24487 & 3.20883 & 1.045 \\
1 & ctio00 & A0620$-$00 & 06:23:15.70 & $-$00:18:20.7 & 209.98050 & $-$6.40597 & 0.292 \\
29 & kpno02 & MADDALENA'SCLOUD & 06:49:24.78 & $-$04:34:07.2 & 216.77023 & $-$2.53092 & 0.944 \\
30 & kpno02 & M1$-$16 & 07:37:14.89 & $-$09:41:07.6 & 226.82573 & 5.59410 & 0.130 \\
59 & kpno04 & OH231.8$+$4.2 & 07:42:01.49 & $-$14:45:18.0 & 231.84038 & 4.14583 & 0.428 \\
2 & ctio00 & PKS0745$-$191 & 07:47:16.70 & $-$19:15:05.0 & 236.37517 & 3.00117 & 0.306 \\
3 & ctio00 & NGC3256 & 10:28:41.70 & $-$43:59:18.3 & 277.54922 & 11.73626 & 0.068 \\
14 & ctio01 & V382VELORUM1999 & 10:44:48.00 & $-$52:18:18.0 & 284.10977 & 5.87716 & 0.258 \\
60 & ctio05 & PSRB1046$-$58 & 10:48:00.00 & $-$58:28:48.0 & 287.37619 & 0.61334 & 0.684 \\
31 & ctio03 & MSH11$-$62 & 11:11:57.38 & $-$60:41:28.0 & 291.05115 & $-$0.13479 & 0.637 \\
15 & ctio01 & NGC3603 & 11:15:49.80 & $-$61:17:23.0 & 291.70792 & $-$0.51893 & 16.683 \\
32 & ctio03 & G292.2$-$0.5 & 11:19:45.94 & $-$61:40:13.5 & 292.28198 & $-$0.70890 & 2.162 \\
33 & ctio03 & CENX$-$3 & 11:20:49.67 & $-$60:37:04.3 & 292.03949 & 0.32333 & 0.665 \\
61 & ctio05 & MYCN18 & 13:39:57.60 & $-$67:20:24.0 & 307.59271 & $-$4.90835 & 0.317 \\
44 & ctio04 & G309.8$+$0.0 & 13:50:12.00 & $-$62:09:36.0 & 309.73349 & $-$0.06672 & 7.336 \\
45 & ctio04 & PSRJ1509$-$5850 & 15:09:36.00 & $-$58:48:18.0 & 320.01083 & $-$0.59235 & 6.358 \\
46 & ctio04 & G322.5$-$0.1 & 15:24:00.00 & $-$57:06:53.2 & 322.52438 & $-$0.16385 & 4.416 \\
16 & ctio01 & 4U1538$-$52 & 15:42:01.40 & $-$52:18:22.0 & 327.42334 & 2.26117 & 1.319 \\
47 & ctio04 & XTEJ1550$-$564 & 15:51:01.07 & $-$56:27:16.6 & 325.90033 & $-$1.81329 & 0.944 \\
62 & ctio05 & ABELL3627 & 16:13:45.60 & $-$60:48:36.0 & 325.24203 & $-$7.03382 & 0.121 \\
34 & ctio03 & 1RXSJ161411.3$-$630657 & 16:14:53.00 & $-$63:12:57.0 & 323.64777 & $-$8.85384 & 0.075 \\
48 & ctio04 & MZ3 & 16:17:09.70 & $-$51:59:30.2 & 331.71680 & $-$1.00867 & 1.132 \\
4 & ctio00 & GROJ1655$-$40 & 16:54:28.90 & $-$39:51:31.7 & 345.02972 & 2.37637 & 0.638 \\
49 & ctio04 & MARS & 17:00:48.53 & $-$26:58:24.2 & 356.04932 & 9.28549 & 0.132 \\
50 & ctio04 & XTEJ1709$-$267 & 17:09:36.71 & $-$26:36:53.0 & 357.51978 & 7.91659 & 0.341 \\
17 & ctio01 & PSRB1706$-$44 & 17:09:42.00 & $-$44:30:45.0 & 343.07489 & $-$2.69993 & 1.319 \\
5 & ctio00 & G347.5$-$0.5a & 17:12:16.10 & $-$39:34:39.6 & 347.33688 & $-$0.16436 & 4.895 \\
6 & ctio00 & G347.5$-$0.5b & 17:15:36.00 & $-$39:58:36.0 & 347.38821 & $-$0.91690 & 1.794 \\
51 & ctio04 & TORNADO & 17:40:00.00 & $-$30:57:59.2 & 357.63300 & $-$0.03307 & 11.271 \\
35 & ctio03 & GC5 & 17:43:04.80 & $-$29:37:48.0 & 359.11887 & 0.10892 & 15.853 \\
36 & ctio03 & GC2 & 17:43:33.12 & $-$29:01:48.0 & 359.68350 & 0.33651 & 4.930 \\
37 & ctio03 & GC3 & 17:45:44.88 & $-$28:25:48.0 & 0.44666 & 0.23978 & 17.838 \\
38 & ctio03 & GC6 & 17:45:49.20 & $-$29:37:48.0 & 359.43039 & $-$0.39852 & 10.212 \\
7 & ctio00 & SGRA* & 17:46:11.20 & $-$28:53:52.0 & 0.09729 & $-$0.08582 & 54.518 \\
39 & ctio03 & GC1 & 17:46:16.32 & $-$29:01:48.0 & 359.99405 & $-$0.17051 & 36.045 \\
8 & ctio00 & SGRB2 & 17:46:43.10 & $-$28:29:42.1 & 0.50199 & 0.02380 & 65.053 \\
9 & ctio00 & GALACTICCENTERARC & 17:47:08.60 & $-$28:53:36.0 & 0.20979 & $-$0.26248 & 20.284 \\
40 & ctio03 & GC4 & 17:48:27.60 & $-$28:25:48.0 & 0.75573 & $-$0.27009 & 30.783 \\
63 & ctio05 & LIMITINGWINDOW & 17:51:48.00 & $-$29:34:12.0 & 0.15160 & $-$1.48171 & 0.704 \\
52 & ctio04 & STANEKWINDOW & 17:54:24.42 & $-$29:49:16.3 & 0.22242 & $-$2.09703 & 0.478 \\
53 & ctio04 & 4U1755$-$33 & 17:58:49.20 & $-$33:51:00.0 & 357.19409 & $-$4.92093 & 0.399 \\
18 & ctio01 & PSRB1757$-$24 & 18:01:48.00 & $-$24:49:39.0 & 5.37027 & $-$1.02458 & 4.030 \\
54 & ctio04 & BAADE'SWINDOW & 18:03:36.00 & $-$29:57:50.7 & 1.08909 & $-$3.89739 & 0.321 \\
55 & ctio04 & G11.4$-$0.1 & 18:11:20.40 & $-$19:14:24.0 & 11.32621 & $-$0.22863 & 8.865 \\
19 & ctio01 & PSR1813$-$36 & 18:16:25.00 & $-$36:20:29.5 & 356.70163 & $-$9.27033 & 0.092 \\
64 & ctio05 & V4641SGR & 18:19:21.38 & $-$25:27:00.0 & 6.73546 & $-$4.80812 & 0.289 \\
56 & ctio04 & PSRB1823$-$13 & 18:26:00.00 & $-$13:37:12.0 & 17.94011 & $-$0.66268 & 7.538 \\
20 & ctio01 & MWC297 & 18:27:59.80 & $-$03:51:02.6 & 26.82325 & 3.44240 & 6.963 \\
41 & ctio03 & GALACTICPLANE & 18:43:27.72 & $-$03:58:12.0 & 28.49013 & $-$0.03869 & 17.281 \\
21 & ctio01 & SGR1900$+$14 & 19:07:38.30 & $+$09:18:08.1 & 43.04858 & 0.66903 & 2.784 \\
65 & ctio05 & 1H1905$+$000 & 19:08:36.00 & $+$00:11:24.0 & 35.06054 & $-$3.73080 & 0.359 \\
10 & kpno00 & B2224$+$65 & 22:25:13.22 & $+$65:32:45.2 & 108.55382 & 6.84190 & 0.460 \\
22 & kpno01 & 3EGJ2227$+$6122 & 22:29:17.00 & $+$61:19:00.9 & 106.70962 & 3.00593 & 1.012 \\
57 & kpno04 & CTB$-$109LOBE & 23:02:18.85 & $+$58:54:40.1 & 109.23905 & $-$1.02858 & 1.032 \\
25 & kpno02 & G116.9$+$0.2 & 23:59:12.53 & $+$62:24:00.0 & 116.92374 & 0.13922 & 0.461 \\ \hline
\end{tabular}
\begin{minipage}{0.7\linewidth}
\tablenotetext{a}{N is the ID of the Mosaic field, which is in
chronological order of the observations.}  
\tablenotetext{b}{The full column N$_{\rm H}$/10$^{22}$, based on
\cite{sch98}.  Thus the N$_{\rm H}$ values are overestimated for most
Galactic plane fields since the Schlegel values for N$_{\rm H}$ are
for the full Galactic absorption whereas optical counterparts for most
sources detected are in the foreground.}
\end{minipage}
\end{center}
\end{table*}

\figmosaicgal

Exposure times were targeted to obtain ``shallow'' exposures to sample
the bright sources in the field and ``deep'' exposures to reach the
survey goal of $\sim$24 mag for 5\% photometry in the R filter and
10\% photometry in the other filters.  This is to ensure that our
primary measurement of H$\alpha -$R, to search for H$\alpha$ bright
optical counterparts, is not compromised by limited sensitivity in the
narrow-band H$\alpha$ filter.  This sensitivity limit is computed
assuming average seeing ($\sim 1.0\arcsec$), airmass ($\sim$1.2), and
lunar phase ($\sim$4 days).  Shallow exposures were 2 seconds for each
of the V, R, I filters and 30 seconds for the H${\alpha}$ filter; deep
exposures were 900 seconds each for V and I, 1200 seconds for R, and
7500 seconds for H${\alpha}$.  Each of the long exposures were divided
into 5 dithered sub-exposures to prevent chip gaps and bad columns in
the data and to provide cosmic-ray rejection.  Stars usually saturate
at $\sim$12 mag in the shallow images and $\sim$17 mag in the deep
ones.  Considering overhead time at the telescope, each field takes
3.5~--~4 hours, allowing us to observe three fields per night.
Absolute minimal observations require the deep R and H${\alpha}$
exposures.

Standard calibration images (biases, dome flats) are taken at the
telescope daily before observing.  Ideal sky flats are usually
constructed from object frames after eliminating all the stars.  In
the Galactic plane, however, the stellar density is too high for this
process, so instead we obtained twilight flats.  Sky flats are
critically important for the I and H$\alpha$ images from the KPNO
4-m to remove a pupil ghost caused by light back-scattered from the
telescope optics that affects the four inner Mosaic CCDs.  Dark images
are not needed for this project as the dark current is very low
($\sim5e^-$/pixel/hr for Mosaic-I and $\sim<2 e^-$/pixel/hr for
Mosaic-II).

\section{Mosaic Data Reduction}
\label{sec:reduction}
The data reduction of Mosaic images is done using the Mosaic Data
Reduction package
(MSCRED)\footnote{http://iraf.noao.edu/iraf/web/irafnews/apr98/irafnews.21.html}
of IRAF\footnote{IRAF (Imaging Reduction and Analysis Facility) is
distributed by the National Optical Astronomy Observatory, which is
operated by the Association of Universities for Research in Astronomy,
Inc., under cooperative agreement with the National Science
Foundation.}.  This section summarizes our reduction process, relying
heavily on the detailed reduction description of Jannuzi et
al.\footnote{\label{Jannuzi}{http://www.noao.edu/noao/noaodeep/ReductionOpt/frames.html}}
for the NOAO Deep Wide Field Survey.

\subsection{CCD reduction}

The raw data are reduced using the {\em IRAF/mscred/ccdproc} package,
including standard CCD corrections.  For the Mosaic-II data, the data
from the two amplifiers of the same CCD are merged.  The Mosaic-I
images are affected by the pupil ghost described in the Mosaic User
Manual\footnote{http://www.noao.edu/kpno/mosaic/manual/man\_sep04.pdf}
and the sky flats are used to remove the ghost at this juncture.  Each
reduced image is then processed to remove cosmic rays and to fix the
bad pixels and columns.

\subsection{Astrometry}
\label{subsec:astrometry}

Astrometry was performed on each image using {\em
IRAF/mscred/msccmatch} which allows for zero-point shifts, scale
changes, and axis rotations.  It also corrects for atmospheric
refraction effects.  Of the two World Coordinate System (WCS) catalogs
accessible by {\em msccmatch} -- the US Naval Observatory A2.0 catalog
(NOAO:USNO-A2)\footnote{http://tdc-www.harvard.edu/software/catalogs/ua2.html}
and the Hubble Space Telescope Guide Star Catalog Version 2
(GSC2@STSCI)\footnote{http://www-gsss.stsci.edu/gsc/gsc2/GSC2home.htm}
-- we chose to use the USNO-A2 catalog because it provides complete
coverage of the sky.  Following coordinate registration, we then
re-project each image to a defined tangent plane using {\em mscimage},
allowing us to stack multiple images.  All re-projections are carried
out using sinc function interpolation because it preserves all spatial
frequencies and is the mathematically correct interpolation method for
well-sampled data.  After the projection, each image needs to have
residual large-scale (field wide) gradients in the sky background
removed.  This is done by using {\em mscskysub} on each projected
image.

\subsection{Image Stacking}
\label{subsec:image_stacking}
Before stacking the images, the relative intensity scales are adjusted
using {\em mscimatch} on each group of images to be stacked.  Finally
the images are stacked using {\em mscstack} with median-combine, which
further removes the cosmic rays, bad pixels and other defects.
Typically we end up with eight final images for each field: shallow,
single images of the V, R, I and H$\alpha$ filters and corresponding
stacked deep images.  These images are available from the \cpl
website$^{\ref{cplweb}}$ (see \citet{gri05}).

\section{Photometry and Photometric Calibration}
\label{sec:photometry}

Photometric analysis uses the standard approach for DAOphot in the
IRAF package {\em noao.digiphot.daophot}
\footnote{http://iraf.noao.edu/docs/photom.html: ``A User's Guide to
Stellar CCD Photometry with IRAF'', P. Massey and L. Davis, 1992; ``A
Reference Guide to the IRAF/DAOPHOT Package'', L. Davis, 1994.}.  The
source lists are generated from all eight images using the task {\em
daofind}, with different threshold, i.e. the detected counts in units
of sky RMS.  For shallow images, the threshold is set high ($\sim$25)
to detect sources brighter than $\sim$19 mag; for the deep images, the
threshold is set low ($\sim$4) to detect all possible sources.  A
comprehensive Master Star List for each field is generated by merging
these eight source lists and removing multiple detections, defined as
sources with positions within 1 pixel (0.26\arcsec) of each other.  So
the source positions in each master star list are on the integer pixel
grid.  Additional duplications will be removed while re-centering
during the PSF fit photometry.  The final star positions are
determined by the center of their PSF.  For each image, 1000 PSF star
candidates were carefully selected, based on their {\em sharpness,
sround} and {\em ground} values from {\em daofind}.  These candidates
should uniformly cover the entire field and are bright enough but not
saturated.  This candidate list is fed into the task {\em psfselect},
which selects the PSF stars.  A Point Spread Function is calculated
for each image from these PSF stars, using {\em psf}, that is constant
over the whole field.  Aperture photometry and PSF fit photometry
(DAOphot) are carried out using tasks {\em phot} and {\em allstar},
respectively, on the Master Star List.

The photometric calibration was obtained from CCD images of \cpl
fields and Landolt standards \citep{lan92} on photometric nights using
the FLWO 1.2-m (north) and CTIO 1.3-m (south) telescopes, so we could
spend all our 4-m time on the \cpl fields.  Typically we take the V,
R, I images of those \cpl fields to be calibrated and 2--3 Landolt
fields (with 5 to a couple dozen Landolt standard stars per field) at
different airmass.  Standard calibration procedures were used ({\em
IRAF/noao.digiphot.photcal}) to compute the photometric
transformations and to determine the V, R, I magnitude standard.  For
the H${\alpha}$ magnitudes, we define the median point of H$\alpha -$R
$\equiv$ 0.

\section{Optical Catalog}
\label{sec:opticalcat}

Even with DAOphot, many false detections (e.g. near the edges, gaps
and saturated stars, under the shadow of bright stars, etc.)  could
still survive the PSF fitting process.  To select the real sources, we
choose objects with the PSF fitting parameter: $sharpness > -1$.
Results with $sharpness \leqslant -1$ are usually too sharp to be real
sources but false detections.  The above selection includes both point
and extended sources.  Point sources usually have $|sharpness| < 1$;
while extended sources (e.g. galaxies) have $sharpness \geqslant 1$.
After selection, for each field, a catalog is established based on the
DAOphot results; each entry in the catalog includes the source ID, RA
and Dec, X and Y position on the image, V, R, I and H$\alpha$
magnitudes and their errors, and the PSF fitting parameters $\chi^2$
and {\em sharpness}.

All the Mosaic optical catalogs will be in the \cpl Online Database
along with the Chandra source optical counterparts, available from the
\cpl website$^{\ref{cplweb}}$ and NOAO Science
Archive\footnote{http://archive.noao.edu/nsa/}.

\section{Astrometric Accuracy}
\label{sec:astrometry}

We examined the astrometric accuracy of our optical catalog using
overlapping areas of the Mosaic fields in our Galactic Center (GC)
mapping.  We matched the positions of identical stars appearing in
different Mosaic fields.  Figure \ref{fig:astrometry} shows the
results from a rectangular (4\arcmin\ in RA and 36\arcmin\ in Dec)
overlapping area between the GC1 and GC2 fields (see Table
\ref{tbl_mosaic}).  The upper left panel shows the offsets of the
identical stars between the two fields; the upper right panel is a
histogram displaying the R magnitude difference of those stars.  The
two lower panels display the histograms of the RA and Dec offsets of
these stars.  The offsets and their errors are a little larger in RA
than in Dec, because the overlapping area in RA is small (matching one
side of one field to the other side of another field) while the
overlapping area in Dec covers the entire Mosaic length.  The standard
deviation of the position mismatch is 0.0984\arcsec\ in RA and
0.0826\arcsec\ in Dec and represents the random precision of
individual stars.  Note that this is the worse-case scenario because
we are comparing the astrometry near the chip edge and at opposite
side of the Mosaic camera.  The astrometry improves toward the center
of the detector.  Even so, the standard deviation is still less than
0.1\arcsec, which is contributed by two Mosaic fields.  Assuming each
one of them has the same contribution to the mismatch, the precision
of each field should be 0.07\arcsec.  To be conservative, we assign
each individual star with an astrometric accuracy of 0.1\arcsec
(1-$\sigma$).  This value is used in Section \ref{sec:chandra_opt} to
identify optical counterparts of Chandra sources.  \figastrometry

\section{H$\alpha$ Emission Objects}
\label{sec:ha}
Accretion-powered binaries are characterized by their (often double
peaked) hydrogen Balmer series emission lines generated in the outer
region of their accretion disks.  Among them, the H$\alpha$ line is
the most prominent.  Therefore we have designed our Mosaic
observations (see \ref{subsec:onservations}) to spend $\sim6\times$
longer exposures in H$\alpha$ than in R to allow maximum sensitivity
to H$\alpha -$R colors (see below).  Since the Mosaic CCDs cover about
5 times more sky than ACIS-I, we expect and found on average 80\% of
the H$\alpha$ emission objects lying outside the ACIS FoV.

To select sources with significant H$\alpha$ excess, we first define
the signal-to-noise ratio, {\em S/N}, in terms of the flux ratio, {\em F},
between H$\alpha$ and R bands as follows.

\begin{eqnarray}
S/N &=& \frac{F - F_0}{\Delta F}       \\
\label{eq_f}
F &=& \left \{ \begin{array}{ll}
          \frac{f_{H\alpha}}{f_R}~     & ~~{\rm if~H\alpha < R ~(H\alpha~emission)} \\
          F_0                          & ~~{\rm if~H\alpha = R}  \\
          \frac{f_R}{f_{H\alpha}}F_0^2 & ~~{\rm if~H\alpha > R ~(H\alpha~absorption)}
	  \end{array}
  \right.
\end{eqnarray}
where H$\alpha$ and R are the magnitudes in the H$\alpha$ and R bands;
$f_{H\alpha}$ and $f_R$ are their fluxes; and $F_0$ corresponds to the
median flux ratio of $\frac{f_{{H\alpha_0}}}{f_{R_0}}$ in the stellar
sample, i.e. when H$\alpha$ = R.\footnote{A factor of $F_0^2$ on the
right hand side of Eq.(\ref{eq_f}) is needed to preserve the continuity of
the function $F$.}  Since by definition $F$ is always greater than or
equal to $F_0$, $S/N$ is always greater than or equal to zero.  The
uncertainty, i.e. noise, of $F$ is $\Delta F$, which by definition is
also always positive.

Since
\begin{eqnarray}
|{\rm H\alpha - R}| &=& 2.5~log\frac{F}{F_0} \\
\Delta|{\rm H\alpha - R}| &=& \frac{2.5}{ln10}~\frac{\Delta F}{F}
\end{eqnarray}
where $\Delta|{\rm H\alpha - R}| = \sqrt{(\Delta H\alpha)^2 + (\Delta
R)^2}$.

Therefore
\begin{eqnarray}
S/N &=& (1-10^{-0.4 |{\rm H\alpha - R}|})\frac{F}{\Delta F} \\
    &=& G \cdot \frac{|{\rm H\alpha - R}|}{\Delta |{\rm H\alpha - R}|}
\end{eqnarray}
where $G = \frac{2.5}{ln10}\cdot\frac{1-10^{-0.4 |{\rm H\alpha - R}|}}{|{\rm
H\alpha - R}|}$ is the so-called $G$ factor, which is a function of
${|{\rm H\alpha - R}|}$ as shown in Figure \ref{fig:snrfactor}.
\figsnrfactor
  
Having defined the {\em S/N}, we select possible H$\alpha$ emission
objects using criteria:
\begin{eqnarray}
\label{eq:cri1}
H\alpha - R &\leqslant& -0.3  \\
\label{eq:cri2}
S/N & \geqslant & 5
\end{eqnarray}
The first criterion (Eq. (\ref{eq:cri1})) selects CVs instead of dMe
stars.  Figure \ref{fig:haew} shows the composite transmission curves
of the Mosaic R and H$\alpha$ filters plus the CCD quantum efficiency,
and H$\alpha-$R as a function of the H$\alpha$ emission equivalent
width (EW), assuming a flat continuum.  By choosing
H$\alpha-$R$\leqslant$$-0.3$ we are selecting objects with H$\alpha$
EW greater than 28\AA.  But the relation between H$\alpha-$R and
H$\alpha$ EW also depends on the spectral type.  Figure
\ref{fig:haewms} shows the H$\alpha-$R as a function of the H$\alpha$
EW for different main sequence stars.\footnote{Based on Kurucz Models
(Dr. R. Kurucz, CD-ROM No. 13, GSFC) from
http://garnet.stsci.edu/STIS/stis\_models.html} Most dMe stars have
EW(H$\alpha$)$<$10\AA\ \citep{moc02}, whereas most CVs (except dwarf
novae in outburst) have EW(H$\alpha$)$>$10\AA\ \citep{wil83, war95,
szk02,szk03,szk04,szk05}.  The criterion of H$\alpha-$R
$\leqslant$$-0.3$ further distances the selection from the large
numbers of dMe stars in each field.  The second criterion
(Eq. (\ref{eq:cri2})) discriminates noise in the faint end.  Because
the G-factor is always less than 1, {\em S/N} is always less than
$\frac{|{\rm H\alpha-R}|}{\Delta |{\rm H\alpha - R}|}$.  Thus using
{\em S/N} instead of $\frac{|{\rm H\alpha-R}|}{\Delta |{\rm H\alpha -
R}|}$ makes the selection more restrictive.  Finally, we visually
examine all the H$\alpha$ objects on the H$\alpha$ and R image pairs
to eliminate false detections.  \fighaew \fighaewms

\section{Chandra Optical Counterparts}
\label{sec:chandra_opt}

\subsection{X-ray source position error}

Serendipitous X-ray sources in the \cpl fields are detected from the
Chandra archival data, using the methods (the ChaMPlane X-ray
processing routines XPIPE and PXP) described in \citet{hon05}.  Each
Chandra source has a position error that depends on the net source
counts and off-axis angle from the aimpoint.  The 95\% confidence
error radii, $r_{x(95\%)}$, are calculated as a function of net counts
and off-axis angle, according to an empirical formula based on the
results of raytrace simulations and XPIPE detections \citep{hon05}.

\subsection{X-ray source boresight correction}

Other than individual random errors, the Chandra sources also have a
small systematic position offset, usually less than 1\arcsec\ relative
to the optical positions.  Before matching X-ray and optical sources,
the X-ray positions are corrected for this boresight difference
($\Delta \alpha$, $\Delta \delta$) that is determined for each
observation separately.

To compute the boresight correction we first select Chandra sources
with well-determined positions, i.e. $r_{x(95\%)}$ smaller than
typically $1\arcsec$ to $2\arcsec$. If in the end it
turns out that the final boresight is based on only a few (2 or 3)
pairs of X-ray/optical matches, the limit on $r_{x(95\%)}$ is increased.
Optical sources are selected from the Optical Catalog to have $R<23$
to guarantee good positions. If no $R$ magnitude is available, we
require that the source is brighter than 23 in either $V$, $I$ or
$H\alpha$ -- in that order of priority.

The resulting X-ray and optical source lists are
cross-correlated. Initially we accept matches inside a large match
radius which combines statistical and systematic errors for a combined
2$\sigma$ value:
\begin{equation}
\label{eq:r0}
R_0 = 2 \times \sqrt{r_x^2 + r_{opt}^2 + r_{xsys}^2}
\end{equation}
\noindent
where \\
$r_x = r_{x(95\%)}/1.95996$, is the 1-$\sigma$ random error on X-ray positions,
assuming the errors follow a Gaussian distribution;\\
$r_{opt} = 0.1\arcsec$ is the random error of individual optical positions
with respect to the astrometric frame (see section \ref{sec:astrometry}); and \\
$r_{xsys} = 0.6\arcsec/1.6449$, is the 1-$\sigma$ typical boresight
offset of X-ray positions ($0.6\arcsec$ is the 90\% absolute accuracy on Chandra
positions\footnote{http://cxc.harvard.edu/cal/docs/cal\_present\_status.html,
http://cxc.harvard.edu/cal/ASPECT/celmon/}).

A weighted (by 1/$r_{x(95\%)}$$^2$) average offset in RA ($\Delta
\alpha$) and Dec ($\Delta \delta$) (defined as the X-ray position
minus the optical position) is computed after excluding X-ray sources
with more than one possible (i.e. within $R_0$) optical
counterpart. Errors on the offsets, $\Delta \alpha_{err}$ and $\Delta
\delta_{err}$, are computed as the square-root of the weighted
variance in $\Delta \alpha$ and $\Delta \delta$ divided by the number
of matches. X-ray/optical matches whose individual offsets lie more
than 3 $\times$ $\Delta \alpha_{err}$ or 3 $\times$ $\Delta
\delta_{err}$ from the average are rejected, and the weighted average
is re-computed.

After correcting the X-ray positions for $\Delta \alpha$ and $\Delta
\delta$, the cross-correlation is repeated. In the second and
consecutive iterations the match-radius is set as follows:
\begin{equation}
\label{eq:r1}
R_1 = 2 \times \sqrt{r_x^2 + r_{opt}^2 + r_{bore}^2}
\end{equation}
\noindent
where $r_{bore}$ is the greater of $\Delta \alpha_{err}$ or $\Delta
\delta_{err}$ in the previous iteration.

The above steps are iterated until the boresight correction converges
to a stable solution, which is reached in typically 2 to 4 passes.
The result is inspected visually against a Mosaic image; optical
sources with positions that look unreliable (e.g.~the position could
be off from the center-of-light, or there could be a nearby undetected
source) are removed from the catalog and the calculation is redone.
The final stable solution of the boresight correction is applied to
all the X-ray positions.  As an example, the boresight solution for
the GRO J0422+32 field is $\Delta\alpha = 0.21\arcsec \pm
0.059\arcsec$ and $\Delta\delta = -0.36\arcsec \pm 0.044\arcsec$.

\subsection{Optical Counterparts}

The Optical Counterparts are found around the boresight corrected
X-ray positions within an error radius of:
\begin{equation}
\label{eq:rerr}
R_{err} = c \times \sqrt{r_x^2 + r_{opt}^2 + r_{bore}^2} 
\end{equation}
where $r_x$ and $r_{opt}$ are defined in Equation \ref{eq:r0};
$r_{bore}$ is the greater of $\Delta \alpha_{err}$ or $\Delta
\delta_{err}$ in the final stable boresight solution; the square root
expression is the combined effective matching error, $\sigma$; $c$ is
a confidence-level scale factor (number of $\sigma$), and typically we
choose $c=3$ to search for optical counterparts.  For each Chandra
source, we search the Mosaic optical catalog (see Section
\ref{sec:opticalcat}) within $R_{err}$ radius for its optical
counterparts.  If there are more than one matches within the $R_{err}$
radius, they are prioritized (for follow-up spectroscopy) by the
matching distance and their H$\alpha-$R color.  This process produces
the Chandra optical counterparts from the Mosaic optical catalogs,
which contain all the objects with magnitude ranging from $\sim$12 to
$\sim$25.  Thus it misses out stars brighter than $\sim$12 magnitude,
which are saturated even in the short Mosaic images.  To recover those
possible missing bright star counterparts, we match all the Chandra
sources, using the $R_{err}$ radii, with the USNO-A2 and GSC2 catalogs
(see Section \ref{subsec:astrometry}).  This process produces all the
Chandra optical counterparts with magnitude brighter than $\sim$19.
When combining the results from these two matches, and removing the
duplications (i.e. objects appear in both matches), we obtain a
complete set of Chandra optical counterparts for a given field.

\section{Results}
\label{sec:results}

\subsection{\cpl Optical Survey Products}

The end products of the \cpl optical photometry for each Mosaic field
are:
\begin{enumerate}
\item An optical catalog of all the objects (regardless of object type
  e.g., stars, CV candidates, QSOs or galaxies, as determined by
  followup spectroscopy) detected in the entire Mosaic field.  Each
  entry includes the source ID, RA and Dec, X and Y position on the
  image, V, R, I and H$\alpha$ magnitudes and their errors, as well as
  the PSF fitting parameters $\chi^2$ and {\em sharpness}, colors
  V$-$R, R$-$I, H$\alpha-$R and their errors, and S/N of H$\alpha-$R
  (as defined in Section \ref{sec:ha}).
\item A list of H$\alpha$ emission sources (as defined in Section
  \ref{sec:ha}) found in the entire Mosaic field.
\item A list or lists of Chandra optical counterparts (as defined in
  Section \ref{sec:chandra_opt}) found for every Chandra observation
  covered by this Mosaic field.
\end{enumerate}

\subsection{Example Results for GRO J0422+32 Field}
\label{subsec:example}

In this section, we use the black hole X-ray nova GRO J0422+32 Chandra
field (ObsID~676, 20~ks ACIS-I observation) as an example to
demonstrate the typical output of the photometry survey.  The optical
catalog from this Mosaic field has 29714 entries, with magnitudes
ranging from 12.4 to 25.5.  The catalog is accessible at the \cpl
website$^{\ref{cplweb}}$.

\begin{table*}
\begin{center}
\caption{\label{tbl:J0422_ha}GRO J0422+32 field: H$\alpha$ emission
  sources within the Full Mosaic field}
\begin{tabular}{rclccccrr} \hline \hline
OptID & RA(J2000) & Dec(J2000) & V(err) & R(err) & I(err) & H$\alpha-$R & S/N & N \\ \hline
187498 & 04 20 17.07 & $+$32 47 20.8 & 21.23(02) & 20.21(01) & 18.86(01) & $-$0.34(02) & 18.92 & 1 \\
\tablenotemark{b}108811 & 04 21 30.28 & $+$33 07 29.2 & 21.05(01) & 20.28(01) & 19.79(02) & $-$0.76(01) & 41.77 & 2 \\
\tablenotemark{*}\tablenotemark{a}97731 & 04 21 42.72 & $+$32 54 27.1 & 21.87(02) & 20.80(01) & 19.84(02) & $-$1.45(01) & 61.43 & 3 \\
99419 & 04 21 40.58 & $+$33 12 04.1 & 21.71(02) & 21.14(01) & 20.77(05) & $-$0.45(02) & 19.18 & 4 \\
37863 & 04 22 43.56 & $+$33 13 32.3 & $-$ & 21.28(03) & 19.55(03) & $-$0.35(04) & 8.30 & 5 \\
226061 & 04 19 53.04 & $+$33 08 28.0 & $-$ & 21.60(03) & 20.77(08) & $-$0.42(04) & 8.56 & 6 \\
170813 & 04 20 33.62 & $+$32 39 48.8 & $-$ & 21.91(03) & 20.19(03) & $-$0.32(04) & 6.58 & 7 \\
\tablenotemark{*}81814 & 04 22 00.25 & $+$32 57 08.0 & 23.34(07) & 21.98(03) & 21.42(08) & $-$0.32(04) & 6.44 & 8 \\
130822 & 04 21 11.70 & $+$32 38 38.4 & 22.98(07) & 22.18(03) & 21.46(09) & $-$0.55(05) & 8.71 & 9 \\
133739 & 04 21 08.55 & $+$33 08 10.4 & 23.16(09) & 22.22(04) & 21.31(07) & $-$0.36(06) & 5.43 & 10 \\
165944 & 04 20 37.67 & $+$33 09 57.1 & 23.16(07) & 22.36(03) & 21.68(09) & $-$0.43(05) & 6.75 & 11 \\
197374 & 04 20 07.98 & $+$32 44 28.8 & 23.17(08) & 22.43(04) & $-$ & $-$1.04(05) & 14.71 & 12 \\
155336 & 04 20 48.59 & $+$32 45 22.0 & 23.21(09) & 22.68(04) & 22.20(16) & $-$0.97(06) & 11.36 & 13 \\
133393 & 04 21 09.44 & $+$32 40 59.9 & 23.57(13) & 22.81(06) & $-$ & $-$0.87(07) & 8.29 & 14 \\
\tablenotemark{*}87347 & 04 21 54.25 & $+$32 47 35.2 & 24.51(24) & 23.21(07) & $-$ & $-$1.54(08) & 10.20 & 15 \\
81911 & 04 22 00.32 & $+$32 42 16.6 & 24.11(16) & 23.22(07) & 22.62(23) & $-$0.76(10) & 5.40 & 16 \\
156015 & 04 20 47.83 & $+$32 50 55.6 & $-$ & 23.29(08) & 22.77(25) & $-$0.89(10) & 6.22 & 17 \\
151985 & 04 20 51.72 & $+$32 45 17.5 & 24.05(20) & 23.63(10) & $-$ & $-$1.80(11) & 8.11 & 18 \\ \hline\hline
\end{tabular}
\begin{minipage}{0.62\linewidth}
\tablenotetext{a}{OptID 97731 is GRO J0422+32.}
\tablenotetext{b}{OptID 108811 is the first CV discovered under the \cpl
  survey, located outside the Chandra ACIS FoV.}
\tablenotetext{*}{Within the ACIS-I FoV.}
\end{minipage}
\vspace{0.15in}
\end{center}
\end{table*}
 
\fighasources
Table \ref{tbl:J0422_ha} is a list of H$\alpha$ emission sources with
  H$\alpha-$R$\leqslant$$-$0.3 and S/N$\geqslant$5 (these criteria
  can, of course, be changed to select different H$\alpha$ sources).
  There are 18 H$\alpha$ sources that satisfy these criteria.  Among
  them, ID~97731 is J0422+32 itself.  It has very strong H$\alpha$
  emission (H$\alpha-$R~=~$-$1.45 and S/N~=~61.4) and therefore it was
  easily detected in our survey.  ID~108811 is the first CV discovered
  under the \cpl project.  It has strong H$\alpha$ emission
  (H$\alpha-$R~=~$-$0.76 and S/N~=~41.8).  It was found outside the
  ACIS FoV so we do not know its X-ray properties.  Its CV status was
  confirmed via spectroscopy \citep{rog05}.  ID~87347 is a very strong
  H$\alpha$ emission object with H$\alpha-$R~=~$-$1.54 and S/N~=~10.2,
  corresponding to EW = 320\AA\ based on Figure \ref{fig:haew}.  It is
  also (barely) detected in B images taken with the Wide Field Camera
  on the 2.5-m Isaac Newton Telescope on La Palma, on Jan 13, 2004. A
  preliminary estimate gives B=25.1(2).  This object is inside the
  ACIS-I FoV.  However, no X-ray emission was detected from this
  object.  The (unabsorbed) flux detection limits (3-$\sigma$) of this
  object are $uFx(0.5-2.0keV)\leqslant 2.5\times 10^{-15}
  ergs\,cm^{-2}sec^{-1}$, $uFx(2.0-8.0keV)\leqslant 9.5\times 10^{-15}
  ergs\,cm^{-2}sec^{-1}$, and $uFx(0.5-8.0keV)\leqslant 5.6\times
  10^{-15} ergs\,cm^{-2}sec^{-1}$, assuming N$_{\rm H}=1.9\times
  10^{21}$, based on \citet{sch98}, and a power-law spectrum with
  $\Gamma = 1.7$.  This yields the absorbed and unabsorbed flux ratios
  $F_x(Hc)/F_r \leqslant 10.4$ and $uF_x(Hc)/uF_r \leqslant 5.2$.
  This object can potentially be a CV, because its EW is much too
  large to be a dMe star.  Deeper Chandra images and optical spectra
  are required, though this shows the generally comparable depths of
  the Mosaic and Chandra-ChaMPlane images.  Figure \ref{fig:hasources}
  shows the R and H$\alpha$ images of these three H$\alpha$ emission
  sources.

Table \ref{tbl:J0422_xp} is a list of {\em point} Mosaic optical
  sources (i.e. $|sharpness| < 1$) found within the match-radii
  (Eq. \ref{eq:rerr}) of all the level 2 Chandra sources\footnote{See
  \citet{hon05} for definition of levels; for this ObsID level 2
  sources are all the valid sources on the ACIS-I chips}. XPIPE
  detected 62 point sources on the four ACIS-I chips.  One of the 62
  is the target -- J0422.  The other 61 are all newly discovered X-ray
  sources, based on a search in the SIMBAD Astronomical
  Database\footnote{SIMBAD Astronomical Database is operated at CDS,
  Strasbourg, France, http://simbad.u-strasbg.fr/cgi-bin/WSimbad.pl}.
  37 of these X-ray sources match with 43 point optical counterparts
  within the 3-$\sigma$ search radius.  Four of the X-ray sources have
  two possible counterparts each and one X-ray source has three
  possible counterparts within the 3-$\sigma$ error circle.
  \begin{table*}
\begin{center}
\caption{\label{tbl:J0422_xp}GRO J0422+32 field: Chandra Optical Counterparts (point sources)}
\begin{tabular}{crccrrcrrrcccrr} \hline \hline
SrcID\tablenotemark{a} & OptID & RA(J2000) & Dec(J2000) & $r$\tablenotemark{b} & $d$\tablenotemark{c} & $\sigma$\tablenotemark{d} & $F_x$~~ & $F_x/F_r$ & $F_x/F_r$ & V(err) & R(err) & I(err) & H$\alpha-$R~ & S/N \\ 
        &       &             &               & (\arcsec) & (\arcsec) & ($\sigma$) & (Bc)\tablenotemark{e} & (Sc)\tablenotemark{f} & (Hc)\tablenotemark{g} & (mag) & (mag) & (mag) & (mag)~ &   \\ \hline
B0\_001 & 80714 & 04 22 01.58 & $+$32 57 29.3 & 0.94 & 0.20 & 0.63 & 43.94 & 2.915 & 5.351 & 21.88(02) & 21.53(02) & 20.72(04) & 0.17(04) & 4.44 \\
B0\_002 & 82718 & 04 21 59.17 & $+$32 57 58.4 & 1.25 & 0.36 & 0.86 & 15.81 & 4.050 & 12.988 & 24.00(17) & 23.20(07) & 22.33(18) & $-$0.04(14) & 0.28 \\
B0\_003 & 89269 & 04 21 51.97 & $+$32 57 06.8 & 0.90 & 0.29 & 0.97 & 14.62 & 3.651 & 10.253 & 23.69(11) & 23.12(06) & 22.28(15) & 0.24(14) & 1.54 \\
B0\_004 & 92541 & 04 21 48.36 & $+$32 58 15.9 & 1.25 & 0.05 & 0.13 & 11.53 & 1.017 & 0.996 & 22.24(03) & 21.69(02) & 21.10(05) & 0.07(04) & 1.95 \\
B0\_005 & \tablenotemark{h}81814 & 04 22 00.25 & $+$32 57 08.0 & 3.06 & 1.15 & 1.13 & 2.71 & 0.296 & 0.402 & 23.34(07) & 21.98(03) & 21.42(08) & $-$0.32(04) & 6.44 \\
B0\_007 & 94647 & 04 21 45.99 & $+$32 58 45.3 & 1.46 & 0.21 & 0.44 & 7.27 & 0.854 & 3.196 & 23.09(07) & 22.42(04) & 21.81(10) & 0.09(07) & 1.19 \\
B0\_008 & 95706 & 04 21 44.84 & $+$33 00 29.2 & 1.52 & 0.35 & 0.70 & 22.68 & 2.692 & 9.002 & 23.03(07) & 22.38(03) & 21.88(12) & 0.14(07) & 1.86 \\
B0\_009 & 70708 & 04 22 12.67 & $+$33 01 26.9 & 5.53 & 0.12 & 0.06 & 13.25 & 0.678 & 0.820 & 21.64(02) & 21.13(01) & 20.67(04) & 0.00(02) & 0.00 \\
B0\_010 & 75612 & 04 22 07.22 & $+$32 57 07.0 & 4.29 & 3.21 & 2.25 & 5.51 & 0.003 & 0.133 & 19.26(01) & 18.48(00) & 17.80(01) & 0.00(01) & 0.42 \\
B0\_010 & 75913 & 04 22 06.87 & $+$32 57 06.2 & 4.29 & 1.56 & 1.09 & 5.51 & 0.046 & 2.372 & 22.83(06) & 21.60(02) & 20.67(04) & $-$0.01(04) & 0.28 \\
B0\_014 & 85213 & 04 21 56.39 & $+$33 03 39.0 & 3.47 & 1.44 & 1.24 & 22.22 & 3.041 & 5.763 & 22.92(07) & 22.32(03) & 22.18(15) & 0.11(07) & 1.52 \\
B0\_015 & 86506 & 04 21 55.04 & $+$33 00 36.3 & 2.94 & 0.24 & 0.24 & 9.08 & 0.201 & 1.471 & 21.47(02) & 20.99(01) & 20.68(04) & 0.08(03) & 3.17 \\
B0\_017 & 72773 & 04 22 10.35 & $+$33 01 26.0 & 5.78 & 0.54 & 0.28 & 7.82 & 0.454 & 0.130 & 21.49(02) & 21.09(01) & 20.61(04) & 0.03(03) & 1.04 \\
B0\_018 & 98282 & 04 21 42.03 & $+$33 02 22.1 & 6.77 & 6.17 & 2.73 & 9.05 & 1.276 & 7.202 & $-$ & 22.83(06) & 22.28(18) & $-$0.04(11) & 0.31 \\
B0\_018 & 98409 & 04 21 41.87 & $+$33 02 23.0 & 6.77 & 4.44 & 1.97 & 9.05 & 0.059 & 0.333 & 20.67(01) & 19.49(01) & 18.35(01) & $-$0.12(01) & 11.63 \\
B0\_018 & 98682 & 04 21 41.59 & $+$33 02 25.8 & 6.77 & 2.13 & 0.94 & 9.05 & 1.522 & 8.595 & 23.57(10) & 23.02(07) & $-$ & 0.19(15) & 1.19 \\
B1\_001 & 87169 & 04 21 54.39 & $+$32 53 09.8 & 1.07 & 0.27 & 0.75 & 7.75 & 0.601 & 0.482 & 22.07(03) & 21.52(02) & 21.16(06) & 0.06(03) & 1.66 \\
B1\_002 & 81268 & 04 22 00.97 & $+$32 52 36.4 & 1.04 & 0.25 & 0.73 & 21.83 & 0.976 & 0.629 & 21.35(02) & 20.89(01) & 20.35(03) & $-$0.01(02) & 0.72 \\
B1\_004 & 92607 & 04 21 48.36 & $+$32 54 04.2 & 1.12 & 0.20 & 0.54 & 2.51 & 0.120 & 1.602 & $-$ & 22.27(03) & 21.29(08) & 0.17(07) & 2.03 \\
B1\_006 & 71512 & 04 22 11.83 & $+$32 56 04.3 & 1.42 & 0.53 & 1.12 & 31.82 & 0.555 & 2.120 & 20.76(01) & 20.36(01) & 19.70(02) & 0.12(02) & 5.22 \\
B2\_003 & 110429 & 04 21 28.65 & $+$32 55 46.9 & 1.14 & 0.24 & 0.64 & 8.04 & 2.817 & $-$~~ & 23.71(12) & 22.96(06) & 21.96(13) & 0.10(12) & 0.76 \\
B2\_004 & 111561 & 04 21 27.42 & $+$32 55 51.5 & 1.64 & 0.60 & 1.09 & 4.63 & 2.067 & 1.541 & $-$ & 23.41(09) & 22.32(17) & 0.04(18) & 0.23 \\
B2\_008 & 123049 & 04 21 17.27 & $+$33 00 31.1 & 3.23 & 0.73 & 0.68 & 21.28 & 1.340 & 4.630 & 22.20(04) & 21.70(03) & $-$ & 0.14(05) & 2.51 \\
B2\_008 & 123477 & 04 21 17.01 & $+$33 00 31.4 & 3.23 & 2.57 & 2.39 & 21.28 & 0.412 & 1.422 & 21.67(02) & 20.42(01) & 19.14(01) & $-$0.08(02) & 4.64 \\
B2\_011 & 122729 & 04 21 17.47 & $+$33 02 05.6 & 8.02 & 0.57 & 0.21 & 11.56 & 0.031 & 0.448 & 19.64(01) & 19.20(01) & 18.53(01) & 0.07(01) & 6.40 \\
B2\_012 & 115387 & 04 21 23.27 & $+$33 02 12.6 & 9.57 & 3.51 & 1.10 & 7.75 & 0.112 & 0.371 & 21.11(01) & 20.08(01) & 19.26(01) & $-$0.06(01) & 5.29 \\
B2\_013 & 133097 & 04 21 09.40 & $+$32 55 42.3 & 8.46 & 2.09 & 0.74 & 6.80 & 0.543 & 13.065 & $-$ & 23.32(09) & 21.70(12) & 0.23(20) & 1.05 \\
B3\_001 & 95243 & 04 21 45.50 & $+$32 51 58.9 & 0.84 & 0.17 & 0.59 & 16.14 & 5.790 & 15.310 & 24.08(16) & 23.49(10) & 22.66(26) & 0.03(21) & 0.14 \\
B3\_003 & \tablenotemark{h}97731 & 04 21 42.72 & $+$32 54 27.1 & 0.70 & 0.10 & 0.43 & 10.15 & 0.422 & 0.249 & 21.87(02) & 20.80(01) & 19.84(02) & $-$1.45(01) & 61.43 \\
B3\_005 & \tablenotemark{i}106618 & 04 21 32.88 & $+$32 53 27.3 & 0.91 & 0.16 & 0.52 & 9.81 & 0.013 & 0.004 & 18.35(00) & 17.00(01) & 15.54(00) & $-$0.19(01) & 24.40 \\
B3\_006 & 95130 & 04 21 45.62 & $+$32 51 14.6 & 0.88 & 0.16 & 0.54 & 25.04 & 2.462 & 13.008 & $-$ & 22.42(04) & 22.01(15) & 0.06(07) & 0.95 \\
B3\_008 & \tablenotemark{i}111135 & 04 21 27.94 & $+$32 53 02.1 & 1.33 & 0.46 & 1.03 & 10.74 & 0.014 & 0.009 & 18.53(01) & 17.08(01) & 15.42(01) & $-$0.20(01) & 15.82 \\
B3\_009 & \tablenotemark{i}111329 & 04 21 27.76 & $+$32 50 38.3 & 1.45 & 0.48 & 0.99 & 15.59 & 1.706 & 2.669 & 22.89(05) & 22.04(02) & 21.59(08) & $-$0.26(04) & 5.26 \\
B3\_010 & 112331 & 04 21 26.61 & $+$32 51 34.8 & 2.38 & 0.75 & 0.95 & 4.45 & 0.002 & 0.001 & 16.52(01) & 15.94(01) & 15.37(01) & $-$0.04(01) & 3.90 \\
B3\_011 & 86630 & 04 21 55.08 & $+$32 47 26.0 & 1.75 & 0.89 & 1.52 & 45.50 & 1.527 & 4.094 & 21.57(02) & 20.93(01) & 20.22(03) & $-$0.01(02) & 0.49 \\
B3\_012 & 88797 & 04 21 52.60 & $+$32 47 01.0 & 7.24 & 6.12 & 2.53 & 6.27 & 0.057 & 0.029 & 19.91(01) & 19.15(01) & 18.45(01) & 0.02(01) & 1.55 \\
B3\_012 & 89291 & 04 21 52.09 & $+$32 47 04.1 & 7.24 & 1.15 & 0.48 & 6.27 & 2.300 & 1.172 & $-$ & 23.16(07) & 22.57(22) & 0.23(17) & 1.23 \\
B3\_013 & 90634 & 04 21 50.62 & $+$32 47 55.1 & 9.63 & 7.25 & 2.26 & 1.18 & 0.041 & 0.297 & 22.40(04) & 21.49(02) & 20.71(04) & $-$0.11(03) & 3.22 \\
B3\_015 & 99529 & 04 21 40.82 & $+$32 49 40.5 & 2.77 & 1.13 & 1.22 & 6.62 & 0.409 & 7.764 & 23.31(09) & 22.85(05) & $-$ & $-$~~~~~ & 0.00 \\
B3\_018 & 115085 & 04 21 23.78 & $+$32 48 37.8 & 1.60 & 0.65 & 1.22 & 43.27 & 2.681 & 6.901 & 22.18(03) & 21.58(02) & 21.08(05) & $-$0.01(04) & 0.28 \\
B3\_019 & 120062 & 04 21 19.83 & $+$32 49 02.6 & 3.15 & 1.14 & 1.08 & 16.01 & 0.014 & 0.005 & 17.78(01) & 16.56(01) & 15.45(01) & $-$0.11(01) & 7.33 \\
B3\_021 & 120888 & 04 21 19.25 & $+$32 47 49.6 & 7.30 & 7.22 & 2.97 & 7.26 & 2.374 & 6.582 & $-$ & 23.41(10) & $-$ & $-$0.02(23) & 0.08 \\
B3\_021 & 121522 & 04 21 18.81 & $+$32 47 48.2 & 7.30 & 1.63 & 0.67 & 7.26 & 0.594 & 1.647 & 22.50(06) & 21.91(03) & $-$ & 0.03(05) & 0.61 \\ \hline \hline
\end{tabular}
\begin{minipage}{.95\linewidth}
\tablenotetext{a}{The full Chandra SrcID has prefix XS00676 for the
  J0422+32 observation (ObsID 676).}

\tablenotetext{b}{$r$ is the 3-$\sigma$ match-radius of the X-ray
source in arcsec.}  

\tablenotetext{c}{$d$ is the matching distance between X-ray
source and its optical counterpart in arcsec.}

\tablenotetext{d}{$\sigma$ is the matching distance in unit of
1-$\sigma$ error radius.}  

\tablenotetext{e}{$F_x(Bc)$ is the absorbed broad band (0.5-8.0 keV)
flux in unit of $10^{-15}ergs\,cm^{-2}sec^{-1}$.  The
unabsorbed-$F_x(Bc) = 1.130 \times F_x(Bc)$.}

\tablenotetext{f}{$F_x(Sc)/F_r$ is the ratio of absorbed soft band flux $F_x$(0.5-2.0 keV) vs. observed
  optical R band flux $F_r$ ($ergs\,cm^{-2}sec^{-1}(1000$\AA$)^{-1}$).  The unabsorbed flux-ratio =
  1.519$\times F_x(Sc)/F_r$}

\tablenotetext{g}{$F_x(Hc)/F_r$ is the ratio of absorbed hard band flux $F_x$(2.0-8.0 keV) vs. $F_r$.  The unabsorbed flux-ratio = 1.017$\times F_x(Hc)/F_r$}

\tablenotetext{h}{Two Chandra sources have strong H$\alpha$ emission
($H\alpha-R \le -0.30$ \& $S/N \ge 5.0$): 97731 is J0422+32; 81814 is a
QSO at z=4.25.}

\tablenotetext{i}{Three Chandra sources have weak H$\alpha$ emission
($-0.30 < H\alpha-R \le -0.19$ and $S/N \ge 5.0$): 106618 and 111135
are dMe stars; 111329 is a QSO at z=1.31.}

\tablenotetext{*}{$F_x(Bc)$, $F_x(Sc)$ and $F_x(Hc)$ are calculated
using a power-law spectral model with $\Gamma = 1.4$.}

\end{minipage}
\end{center}

\end{table*}

Table \ref{tbl:J0422_xg} is a list of {\em extended} optical
  counterparts (i.e. $sharpness \geqslant 1$) of Chandra sources from
  the Mosaic catalog.  There are 3 X-ray sources which match with 3
  extended optical counterparts.  They are most likely galaxies.

A search of USNO-A2 and GSC2 catalogs yields 6 and 5 optical
  counterparts, respectively.  However, all of them are duplicates,
  i.e. they are already included in the list of the Mosaic optical
  counterparts.  Therefore, by combining Table \ref{tbl:J0422_xp} and
  Table \ref{tbl:J0422_xg}, there are 40 Chandra sources matching with
  46 (point or extended) optical counterparts, which is the complete
  set of Chandra optical counterparts in the J0422+32 field.
  \begin{table*}
\begin{center}
\caption{\label{tbl:J0422_xg}GRO J0422+32 field: Chandra Optical Counterparts (extended sources)}
\begin{tabular}{crccrrcrccccccr} \hline \hline
SrcID\tablenotemark{a} & OptID & RA(J2000) & Dec(J2000) & $r$\tablenotemark{b} & $d$\tablenotemark{c} & $\sigma$\tablenotemark{d} & $F_x$~~ & $F_x/F_r$ & $F_x/F_r$ & V(err) & R(err) & I(err) & H$\alpha-$R~ & S/N \\ 
        &       &             &               & (\arcsec) & (\arcsec) & ($\sigma$) & (Bc)\tablenotemark{e} & (Sc)\tablenotemark{f} & (Hc)\tablenotemark{g} & (mag) & (mag) & (mag) & (mag)~ &   \\ \hline
B1\_005 & 71306 & 04 22 12.08 & $+$32 53 57.4 & 2.45 & 0.36 & 0.45 & 10.01 & $-$ & $-$ & 19.14(04) & $-$ & 17.38(04) & $-$ & 0.00 \\
B1\_009 & 81365 & 04 22 00.85 & $+$32 51 04.0 & 3.29 & 0.95 & 0.86 & 7.29 & 0.382 & 1.013 & 22.71(06) & 21.41(03) & 20.31(04) & 0.05(05) & 1.01 \\
B2\_001 & 105760 & 04 21 33.83 & $+$32 55 57.4 & 0.70 & 0.32 & 1.36 & 31.96 & 1.018 & 4.109 & 21.88(05) & 21.05(06) & 20.08(05) & 0.18(09) & 1.89 \\ \hline \hline
\end{tabular}
\begin{minipage}{0.90\linewidth}
See footnotes of Table \ref{tbl:J0422_xp} for column definitions.
\end{minipage}
\end{center}
\end{table*}

Table \ref{tbl:J0422_noxm} is a list of 22 Chandra sources without
 optical counterparts, with their optical magnitude limit as measured
 with the Mosaic photometry.  \begin{table*}
\begin{center}
\caption{\label{tbl:J0422_noxm}GRO J0422+32 field: Chandra Sources without Optical Counterparts}
%\vspace{0.1in}
%\footnotesize
%\scriptsize
%\small
\begin{tabular}{cccrrrcccc} \hline \hline
\multicolumn{6}{c}{} & \multicolumn{4}{c}{Magnitude limits\tablenotemark{e}} \\
SrcID\tablenotemark{a} & RA(J2000) & Dec(J2000) & $F_x$~~ & $F_x$~~ & $F_x$~~ & V & R & I & H  \\ 
            &             &         & (Bc)\tablenotemark{b} &
        (Sc)\tablenotemark{c} & (Hc)\tablenotemark{d} & (mag) & (mag) & (mag) & (mag) \\ \hline
B0\_006 & 04 21 49.07 & $+$32 58 46.1 & 4.68 & 1.07 & 3.53 & 24.80 & 24.70 & 22.90 & 24.30 \\
B0\_011 & 04 21 59.30 & $+$33 01 01.2 & 10.32 & 2.11 & 9.03 & 24.71 & 24.96 & 22.86 & 24.44 \\
B0\_012 & 04 21 59.34 & $+$33 00 21.3 & 4.36 & 1.67 & 0.10 & 24.63 & 25.03 & 23.01 & 24.29 \\
B0\_013 & 04 21 56.36 & $+$33 03 05.7 & 13.30 & 1.20 & 18.96 & 24.61 & 24.63 & 22.97 & 24.22 \\
B0\_016 & 04 21 40.52 & $+$33 03 08.3 & 14.47 & 2.39 & 15.51 & 24.69 & 24.96 & 22.98 & 24.16 \\
B1\_003 & 04 21 52.74 & $+$32 53 45.5 & 7.59 & 2.27 & 3.12 & 24.48 & 24.28 & 22.99 & 24.08 \\
B1\_007 & 04 22 01.71 & $+$32 51 57.5 & 3.94 & 1.04 & 2.25 & 24.65 & 24.85 & 23.00 & 24.33 \\
B1\_008 & 04 22 13.56 & $+$32 51 45.4 & 6.29 & 1.24 & 5.65 & 24.55 & 24.93 & 22.94 & 24.13 \\
B2\_002 & 04 21 40.39 & $+$32 55 57.9 & 6.50 & 0.98 & 7.28 & 24.69 & 24.72 & 22.88 & 24.30 \\
B2\_005 & 04 21 38.25 & $+$32 58 40.2 & 9.76 & 2.31 & 7.03 & 24.82 & 24.78 & 22.78 & 24.38 \\
B2\_006 & 04 21 28.75 & $+$32 56 58.2 & 4.01 & $-$~ & 7.53 & 24.72 & 24.66 & 22.84 & 24.30 \\
B2\_007 & 04 21 22.15 & $+$33 01 07.0 & 11.22 & 0.89 & 16.66 & 24.58 & 25.01 & 22.87 & 24.16 \\
B2\_009 & 04 21 29.63 & $+$33 01 10.8 & 18.15 & 4.71 & 11.26 & 24.76 & 24.97 & 22.96 & 24.43 \\
B2\_010 & 04 21 25.01 & $+$32 59 43.0 & 4.14 & 0.48 & 5.43 & 24.89 & 24.95 & 22.87 & 24.18 \\
B2\_014 & 04 21 19.28 & $+$32 58 07.2 & 7.03 & 2.19 & 2.58 & 24.55 & 24.67 & 22.90 & 24.22 \\
B3\_002 & 04 21 44.82 & $+$32 54 04.1 & 7.67 & 1.73 & 5.85 & 24.70 & 24.94 & 22.84 & 24.44 \\
B3\_004 & 04 21 41.18 & $+$32 53 13.5 & 14.85 & 3.97 & 8.29 & 24.59 & 24.90 & 22.99 & 24.16 \\
B3\_007 & 04 21 43.45 & $+$32 49 44.7 & 8.79 & 2.10 & 6.03 & 24.58 & 25.03 & 23.10 & 24.33 \\
B3\_014 & 04 21 46.60 & $+$32 52 40.4 & 1.92 & 0.50 & 1.14 & 24.67 & 24.73 & 22.85 & 24.20 \\
B3\_016 & 04 21 26.98 & $+$32 48 58.9 & 5.06 & 1.45 & 2.31 & 24.61 & 24.88 & 23.04 & 24.09 \\
B3\_017 & 04 21 25.51 & $+$32 52 28.8 & 2.90 & 1.15 & $-$~ & 24.76 & 24.84 & 22.97 & 24.40 \\
B3\_020 & 04 21 18.66 & $+$32 52 55.6 & 6.19 & 1.79 & 2.79 & 24.65 & 24.73 & 22.91 & 24.25 \\ \hline \hline
\end{tabular}
\begin{minipage}{.65\linewidth}
\tablenotetext{a}{The full Chandra SrcID has prefix XS00676 for the J0422+32 observation (ObsID 676).}
\tablenotetext{b}{$F_x(Bc)$ is the absorbed broad band (0.5-8.0 keV) flux in unit of $10^{-15}ergs\,cm^{-2}sec^{-1}$.}  
\tablenotetext{c}{$F_x(Sc)$ is the absorbed soft band (0.5-2.0 keV) flux in unit of $10^{-15}ergs\,cm^{-2}sec^{-1}$.}
\tablenotetext{d}{$F_x(Hc)$ is the absorbed hard band (0.5-8.0 keV) flux in unit of $10^{-15}ergs\,cm^{-2}sec^{-1}$.}
\tablenotetext{e}{Magnitude limit is 5-$\sigma$ above the sky RMS.}
\end{minipage}
\end{center}

\end{table*}

Table \ref{tbl:J0422_bs} is a list of 5 bright stars in the Henry
 Draper (HD) Catalog, found within the ACIS-I FoV through a SIMBAD
 search, with their known spectral type and (for late-type stars only)
 expected X-ray flux ranges for late-type stars only based on ROSAT
 observations \citep{sch04}.  None of these 5 stars were detected by
 the Chandra observation of J0422.  The soft band flux detection limit
 at position of these 5 stars are also listed in the Table.
 \begin{table*}
\begin{center}
\caption{\label{tbl:J0422_bs}J0422+32 field: Bright HD Stars
  within the ACIS-I FoV}
\begin{tabular}{cccrccrcccc} \hline\hline
ID\tablenotemark{a} & RA(J2000) & Dec(J2000) & B~ & V & Type & d\tablenotemark{b} & N$_{\rm H}/10^{22}$\tablenotemark{c} & 
Expected $F_x$ & $F_x$-lim & $uF_x$-lim \\
&  & &  & & & (pc) & $cm^{-2}$ & Min -- Max\tablenotemark{d} & (Sc)\tablenotemark{e} & (Sc)\tablenotemark{f} \\ \hline
HD281972 & 04:21:12.70 & $+$32:54:39.2 & 9.5 & $-$ & B9 & 748 & 0.189 &  -- & 1.63 & 2.59 \\
HD281969 & 04:21:51.45 & $+$32:53:34.2 & 11.7 & 11.0 & A2 & 344 & 0.194 & -- & 1.56 & 2.50 \\
HD281970 & 04:21:46.14 & $+$32:57:31.5 & 11.6 & 10.9 & F2 & 183 & 0.188 & 0.08 -- 134 & 2.11 & 3.34 \\
HD281973 & 04:21:11.25 & $+$32:49:48.8 & 10.07 & 9.55 & F8 & 129 & 0.200 & 0.16 -- 270 & 3.85 & 6.24 \\
HD281971 & 04:21:19.78 & $+$33:00:17.4 & 9.5 & $-$ & K0 & 36 & 0.174 & 2.12 -- 789 & 2.82 & 4.33 \\ \hline\hline
\end{tabular}
\begin{minipage}{0.73\linewidth}
\tablenotetext{a}{ID from Henry Draper (HD) catalog.}
\tablenotetext{b}{d is the distance in parsec, estimated from V,
M$_{\rm V}$ (from spectral type), and E(B-V), assuming the object is a
main sequence star.}  \tablenotetext{c}{The full column N$_{\rm H}/10^{22}$,
based on \citet{sch98}.}  \tablenotetext{d}{The expected X-ray flux
range in the ROSAT band (0.1$-$2.4 keV), in unit of
$10^{-15}ergs\,cm^{-2}sec^{-1}$, assuming main sequence stars, based
on \citet{sch04}.  The expected X-ray flux from giants are even
lower.}  \tablenotetext{e}{$F_x$-lim is the absorbed soft band
($0.5-2.0$ keV) flux detection limit (3-$\sigma$) of this Chandra observation, in
unit of $10^{-15}ergs\,cm^{-2}sec^{-1}$.}
\tablenotetext{f}{$uF_x$-lim is the unabsorbed soft band flux
detection limit (3-$\sigma$).}
\end{minipage}

\end{center}
\end{table*}

Figure \ref{fig:J0422_x_o_id_ha} shows the final stacked deep R image
of the field GRO J0422+32, which consists of 10 individual (two sets
of 5 dithered) images, with a total exposure time of 2100~seconds.  It
also shows the ACIS, Chandra sources and their optical counterparts
and H$\alpha$ emission sources overlay.  Figure
\ref{fig:J0422_x_o_id_aim_ha} is the same figure but zoomed in around
the ACIS-I aimpoint.

Figure \ref{fig:J0422_cmd_xm_acis} shows the J0422+32 field (H$\alpha
-$R) vs. R color-magnitude and (V$-$R) vs. (R$-$I) color-color
diagrams within the ACIS-I FoV.  Figure \ref{fig:J0422_cmd_xm_out} shows
the same diagrams of objects outside the ACIS-I FoV.

%\clearpage
\figJxoidha \figJxoidaimha \figJcmdxmacis \figJcmdxmout

\subsection{Follow-up Observations}

The next step of the \cpl optical survey is to obtain the optical
spectra of all the H$\alpha$ emission sources and the Chandra optical
counterparts found in the \cpl photometry in order to determine the
nature of the objects.  This is the final, crucial step for completing
the survey.  We have been conducting the spectroscopic follow-up using
the WIYN 3.5-m \citep{rog05} and MMT 6.5-m telescopes for the northern
\cpl fields and the CTIO 4-m and Magellan 6.5-m telescopes for the
southern fields.  Because of the severe extinction in visible bands
towards the Galactic center, we are also carrying out infrared imaging
photometry for the Galactic bulge fields \citep{lay05}.

\section{Summary}

We have successfully conducted and completed the optical imaging and
Mosaic photometry for the ChaMPlane survey.  Considerable followup
work (spectroscopy as well as additional photometry analysis) is now
in progress to identify the nature of the sources and will be reported
in subsequent papers. The photometry survey obtained 65 Mosaic fields,
or $\sim$23 square degrees in the Galactic plane and covers 154
Chandra observations on 105 distinct Chandra fields, which is what we
proposed to accomplish.  Using 6 Mosaic pointings, we mapped out 2.2
square degrees around the Galactic center using V, R, I and H$\alpha$
filters to cover 58 Chandra ACIS observations.  This is the deepest
optical survey towards the Galactic center, so far.

This paper summarizes our Mosaic photometry observations and describes
our data reduction method.  Deep Mosaic imaging produces comprehensive
optical catalogs for each \cpl field.  The R and H$\alpha$
differential photometry efficiently detects H$\alpha$ emission
sources.  Our search method effectively finds the optical counterparts
for the majority of Chandra sources in the low extinction fields.
Spectroscopy follow-ups for the Chandra optical counterparts and
H$\alpha$ emission sources for classification complete the \cpl
survey. All the optical catalogs produced by the \cpl optical survey
will be available at the \cpl Online Database and NOAO Science
Archive.  This legacy Optical Database will provide a rich resource
for Galactic astronomy.

\acknowledgments

We would like to thank following people for participating in varies
stages of this project: P.D. Edmonds, J.E. McClintock, M.R. Garcia,
R. Cameron, A. Cool, H. Cohn, P. Lugger, A, Rogel, S. Slavin,
D. Hoard, and S. Wachter. 

This work is supported in part by NASA/Chandra grants AR1-2001X,
AR2-3002A, AR3-4002A, AR4-5003A, NSF grant AST-0098683 and the Chandra
X-ray Center.  We thank NOAO for its support via the Long Term Survey
program.

Facilities: \facility{CXO(ACIS)}, \facility{CTIO 4m(Mosaic)}, \facility{KPNO 4m(Mosaic)}.

\end{document}